\documentclass[reprint,showkeys,longbibliography,nofootinbib,superscriptaddress,aps,notitlepage]{revtex4-2}%

\usepackage[utf8]{inputenc}
\usepackage[dvipsnames]{xcolor}
\usepackage{mathtools}
\usepackage{amsmath}
\usepackage{amssymb}
\usepackage[colorlinks]{hyperref}
\usepackage{graphicx}
\usepackage{multirow}
\usepackage{physics}
\usepackage{siunitx}
\usepackage{xspace} 
\usepackage{lineno}
\usepackage{dcolumn}

\newcommand\myshade{80}
\colorlet{mylinkcolor}{ForestGreen}
\colorlet{mycitecolor}{Red}
\colorlet{myurlcolor}{violet}

\hypersetup{
  linkcolor  = mylinkcolor!\myshade!black,
  citecolor  = mycitecolor!\myshade!black,
  urlcolor   = myurlcolor!\myshade!black,
  colorlinks = true
}

\newcommand{\MeV}{~MeV/c$^{2}$}

\newcommand{\qcdark}{\texttt{QCDark}}
\newcommand{\qcdarktwo}{\texttt{QCDark2}}

\newcommand{\damascussun}{\texttt{DaMaSCUS-SUN}}

\definecolor{myDarkBlue}{RGB}{0, 0, 139}

\usepackage[normalem]{ulem}

\usepackage{cuted}
\usepackage{siunitx}
\DeclareSIUnit\clight{\text{\ensuremath{c}}}
\DeclareSIUnit\MeVpercsq{\mega\eV\per\clight\squared}
\DeclareSIUnit\micra{\micro\meter}
\DeclareSIUnit\micrasq{\micro\meter\squared}
\newcommand{\electron}{\textit{e}$^{-}$}

\begin{document}
\preprint{}

\title{Constraints on Sub-MeV Dark Matter from Solar Reflection with DAMIC-M}

\author{K.\,Aggarwal}
\affiliation{Center for Experimental Nuclear Physics and Astrophysics, University of Washington, Seattle, WA, United States}

\author{I.\,Arnquist}
\affiliation{Pacific Northwest National Laboratory (PNNL), Richland, WA, United States} 

\author{N.\,Avalos}
\affiliation{Laboratoire de physique nucl\'{e}aire et des hautes \'{e}nergies (LPNHE), Sorbonne Universit\'{e}, Universit\'{e} Paris Cit\'{e}, CNRS/IN2P3, Paris, France}

\author{X.\,Bertou}
\affiliation{CNRS/IN2P3, IJCLab, Universit\'{e} Paris-Saclay, Orsay, France}
\affiliation{Laboratoire de physique nucl\'{e}aire et des hautes \'{e}nergies (LPNHE), Sorbonne Universit\'{e}, Universit\'{e} Paris Cit\'{e}, CNRS/IN2P3, Paris, France}

\author{N.\,Castell\'{o}-Mor}
\affiliation{Instituto de F\'{i}sica de Cantabria (IFCA), CSIC - Universidad de Cantabria, Santander, Spain}

\author{C.\,Centeno-Lorca}
\affiliation{Instituto de F\'{i}sica de Cantabria (IFCA), CSIC - Universidad de Cantabria, Santander, Spain}

\author{A.E.\,Chavarria}
\affiliation{Center for Experimental Nuclear Physics and Astrophysics, University of Washington, Seattle, WA, United States}

\author{A.R. \,Chriss}
\affiliation{Kavli Institute for Cosmological Physics and The Enrico Fermi Institute, The University of Chicago, Chicago, IL, United States} 

\author{J.\,Cuevas-Zepeda}
\affiliation{Kavli Institute for Cosmological Physics and The Enrico Fermi Institute, The University of Chicago, Chicago, IL, United States}

\author{A.\,Dastgheibi-Fard}
\affiliation{LPSC LSM, CNRS/IN2P3, Universit\'{e} Grenoble-Alpes, Grenoble, France}

\author{C.\,De Dominicis}
\affiliation{Laboratoire de physique nucl\'{e}aire et des hautes \'{e}nergies (LPNHE), Sorbonne Universit\'{e}, Universit\'{e} Paris Cit\'{e}, CNRS/IN2P3, Paris, France}

\author{O.\,Deligny}
\affiliation{CNRS/IN2P3, IJCLab, Universit\'{e} Paris-Saclay, Orsay, France}

\author{J.\,Duarte-Campderros}
\affiliation{Instituto de F\'{i}sica de Cantabria (IFCA), CSIC - Universidad de Cantabria, Santander, Spain}

\author{E.\,Estrada}
\affiliation{Centro At\'{o}mico Bariloche and Instituto Balseiro, Comisi\'{o}n Nacional de Energ\'{i}a At\'{o}mica (CNEA), Consejo Nacional de Investigaciones Cient\'{i}ficas y T\'{e}cnicas (CONICET), Universidad Nacional de Cuyo (UNCUYO), San Carlos de Bariloche, Argentina}

\author{R.\,Ga\"{i}or}
\affiliation{Laboratoire de physique nucl\'{e}aire et des hautes \'{e}nergies (LPNHE), Sorbonne Universit\'{e}, Universit\'{e} Paris Cit\'{e}, CNRS/IN2P3, Paris, France}

\author{E.-L.~Gkougkousis}
\affiliation{Universit\"{a}t Z\"{u}rich Physik Institut, Z\"{u}rich, Switzerland}

\author{T.\,Hossbach}
\affiliation{Pacific Northwest National Laboratory (PNNL), Richland, WA, United States} 

\author{L.\,Iddir}
\affiliation{Laboratoire de physique nucl\'{e}aire et des hautes \'{e}nergies (LPNHE), Sorbonne Universit\'{e}, Universit\'{e} Paris Cit\'{e}, CNRS/IN2P3, Paris, France}

\author{B.~J.~Kavanagh}
\affiliation{Instituto de F\'{i}sica de Cantabria (IFCA), CSIC - Universidad de Cantabria, Santander, Spain}

\author{B.\,Kilminster}
\affiliation{Universit\"{a}t Z\"{u}rich Physik Institut, Z\"{u}rich, Switzerland}


\author{I.\,Lawson}
\affiliation{SNOLAB, Lively, ON, Canada }

\author{A.\,Letessier-Selvon}
\affiliation{Laboratoire de physique nucl\'{e}aire et des hautes \'{e}nergies (LPNHE), Sorbonne Universit\'{e}, Universit\'{e} Paris Cit\'{e}, CNRS/IN2P3, Paris, France}

\author{H.\,Lin}
\affiliation{Department of Physics and Astronomy, Johns Hopkins University, Baltimore, MD, United States}

\author{P.\,Loaiza}
\affiliation{CNRS/IN2P3, IJCLab, Universit\'{e} Paris-Saclay, Orsay, France}

\author{A.\,Lopez-Virto}
\affiliation{Instituto de F\'{i}sica de Cantabria (IFCA), CSIC - Universidad de Cantabria, Santander, Spain}

\author{R.\,Lou}
\affiliation{Kavli Institute for Cosmological Physics and The Enrico Fermi Institute, The University of Chicago, Chicago, IL, United States}

\author{H. Lumengo-Kidimbu}
\affiliation{Laboratoire de physique nucl\'{e}aire et des hautes \'{e}nergies (LPNHE), Sorbonne Universit\'{e}, Universit\'{e} Paris Cit\'{e}, CNRS/IN2P3, Paris, France}


\author{S.\,Munagavalasa}
\affiliation{Kavli Institute for Cosmological Physics and The Enrico Fermi Institute, The University of Chicago, Chicago, IL, United States}

\author{J.\,Noonan}
\affiliation{Kavli Institute for Cosmological Physics and The Enrico Fermi Institute, The University of Chicago, Chicago, IL, United States}

\author{D.\,Norcini}
\affiliation{Department of Physics and Astronomy, Johns Hopkins University, Baltimore, MD, United States}

\author{S.\,Paul}
\affiliation{Kavli Institute for Cosmological Physics and The Enrico Fermi Institute, The University of Chicago, Chicago, IL, United States}

\author{P.\,P\'{e}rez-Cobo}
\affiliation{Instituto de F\'{i}sica de Cantabria (IFCA), CSIC - Universidad de Cantabria, Santander, Spain}

\author{P.\,Privitera}
\affiliation{Kavli Institute for Cosmological Physics and The Enrico Fermi Institute, The University of Chicago, Chicago, IL, United States}
\affiliation{Laboratoire de physique nucl\'{e}aire et des hautes \'{e}nergies (LPNHE), Sorbonne Universit\'{e}, Universit\'{e} Paris Cit\'{e}, CNRS/IN2P3, Paris, France}

\author{P.\,Robmann}
\affiliation{Universit\"{a}t Z\"{u}rich Physik Institut, Z\"{u}rich, Switzerland}

\author{B.\,Roach}
\affiliation{Kavli Institute for Cosmological Physics and The Enrico Fermi Institute, The University of Chicago, Chicago, IL, United States}

\author{D.\,Rosenmerkel}
\affiliation{Department of Physics and Astronomy, Johns Hopkins University, Baltimore, MD, United States}

\author{M.\,Settimo}
\affiliation{SUBATECH, Nantes Universit\'{e}, IMT Atlantique, CNRS/IN2P3, Nantes, France}

\author{R.\,Smida}
\affiliation{Kavli Institute for Cosmological Physics and The Enrico Fermi Institute, The University of Chicago, Chicago, IL, United States}

\author{M.\,Traina}
\affiliation{Instituto de F\'{i}sica de Cantabria (IFCA), CSIC - Universidad de Cantabria, Santander, Spain}

\author{R.\,Vilar}
\affiliation{Instituto de F\'{i}sica de Cantabria (IFCA), CSIC - Universidad de Cantabria, Santander, Spain}

\author{R.\,Yajur}
\affiliation{Kavli Institute for Cosmological Physics and The Enrico Fermi Institute, The University of Chicago, Chicago, IL, United States}

\author{D.\,Venegas-Vargas}
\affiliation{Department of Physics and Astronomy, Johns Hopkins University, Baltimore, MD, United States}

\author{C.\,Zhu}
\affiliation{Department of Physics and Astronomy, Johns Hopkins University, Baltimore, MD, United States}

\author{Y.\,Zhu}
\affiliation{Laboratoire de physique nucl\'{e}aire et des hautes \'{e}nergies (LPNHE), Sorbonne Universit\'{e}, Universit\'{e} Paris Cit\'{e}, CNRS/IN2P3, Paris, France}

\collaboration{DAMIC-M Collaboration}

\date{\today} 

\begin{abstract}
\noindent  
The Sun acts as a natural dark matter accelerator. Galactic halo particles scattering in the solar plasma emerge with velocities well beyond the Galactic escape speed, providing a boosted flux that extends the kinematic reach of direct detection experiments into the sub-MeV mass regime. We present constraints on solar-reflected dark matter (SRDM) from the DAMIC-M prototype detector using $\sim$1.3~kg-day of data acquired with silicon skipper charge coupled devices (CCDs) at the Modane Underground Laboratory. Exploiting the spatial diffusion signature of low-energy electron recoils, we derive 90\% C.L. upper limits on the DM-electron scattering cross section for both heavy- and ultralight-mediator benchmarks, reaching $\bar{\sigma}_e\sim 3.16 \cdot 10^{-37} $\unit{\cm\squared} at $0.1$ MeV. For the ultralight mediator, our limits are competitive with the world-leading constraints, achieved with an integrated exposure of $\sim1.3$~kg-day.
These results probe the parameter space between stellar-cooling bounds and terrestrial limits from standard halo searches, a region inaccessible to direct detection experiments relying solely on the standard halo flux. 
\end{abstract}

\maketitle


\section{Introduction} 
\label{sec:introduction}
Astrophysical and cosmological observations provide compelling evidence for dark matter (DM), a non-luminous component that dominates the matter content of the Universe~\cite{Bertone:2004pz}. Despite extensive experimental efforts, however, its microscopic nature remains unknown~\cite{Gaskins:2016cha,Boveia:2018yeb,Billard:2021uyg}. A broad class of well-motivated models predicts light, sub-GeV DM particles in a hidden sector that couples feebly to the Standard Model, for instance through a dark-photon mediator~\cite{Boehm:2003ha,Pospelov:2007mp,Hooper:2008im,Chu:2011be,Knapen:2017xzo}. In this mass range, conventional searches based on nuclear recoils rapidly lose sensitivity~\cite{Drukier:1983gj,Goodman:1984dc,Drukier:1986tm}, whereas DM--electron scattering remains accessible in low-threshold semiconductor detectors~\cite{Essig:2011nj,Essig:2015cda}. Silicon skipper charge-coupled devices (CCDs) are particularly well suited to this program because they combine sub-electron charge resolution with extremely low dark current~\cite{DAMIC:2011khz,DAMICsnolab,SENSEI:2023zdf}. Using this technology, DAMIC-M has recently pushed direct-detection sensitivity to the leading edge over a broad sub-GeV DM mass range~\cite{DAMIC-M:2023gxo,DAMIC-M:2023DM}, excluding for the first time part of the ultralight-mediator freeze-in benchmark for hidden-sector dark matter~\cite{DAMIC-M:2025}.

At low mass, the reach of halo-based searches is strongly constrained by the kinematics of the incident DM population. In the Standard Halo Model (SHM)~\cite{Green:2017odb}, the DM speed distribution in the laboratory frame is bounded by the Galactic escape speed and the motion of the Earth, restricting the maximum energy available for electron ionization. As the DM mass decreases, this kinematic restriction suppresses the ionization rate and eventually closes the phase space available for detectable electron ionization. DAMIC-M has already exploited the time dependence induced by Earth-scattering of the DM halo flux (see e.g.~\cite{Kouvaris:2014lpa,Kavanagh:2016pyr,Lantero-Barreda:2025sgy}) to extend its sensitivity to lower masses through a search for daily modulation~\cite{DAMIC-M:2023DM,DAMIC-M:2025ltz}. A complementary approach is to search for a halo component that has been accelerated before reaching the detector~\cite{Bringmann:2018cvk,Ema:2018bih,Wang:2021jic,Granelli:2022ysi,Cappiello:2022exa}. Such boosted components can arise, for example, from scattering with cosmic rays or from reflection in the Sun. In this work we focus on the latter mechanism, solarreflection~\cite{An:2017ojc,Emken:2017hnp,DaMaSCUS-SUN:2022,DaMaSCUS-SUN:2024}, for which the flux prediction and solar propagation effects are treated explicitly. A dedicated interpretation in terms of cosmic-ray-reflected DM requires a separate signal calculation and is left for future work. In the solar reflection, Galactic-halo DM scatters in the solar medium and emerges with speeds significantly larger than those of the virialized halo population, thereby opening otherwise inaccessible phase space for DM--electron scattering. This mechanism enables direct-detection experiments to probe low-mass parameter space that is inaccessible to conventional SHM-based searches~\cite{SENSEI:2023zdf,CDEX:2023wfz,XENON:2025dpl}.

In this work, we search for solar-reflected dark matter (SRDM) using data from the DAMIC-M Low Background Chamber (LBC). This search uses the `pattern analysis' strategy previously used by the DAMIC-M search in Ref.~\cite{DAMIC-M:2025}, which exploits the spatial information provided by skipper CCDs in the few-electron regime. Specifically, the binned CCD readout configuration allows few-electron events broadened by diffusion to be classified as one-dimensional clusters of adjacent charged pixels, referred to as charge patterns. The pattern analysis allows few-electron events -- such as those due to DM-induced ionization -- to be statistically discriminated from dark current events, which typically appear as isolated charged pixels. We use the same selected LBC data set and pattern-count observables as in Ref.~\cite{DAMIC-M:2025}, and reinterpret them in terms of a solar-reflected population of sub-GeV DM boosted by scattering in the Sun. 

In Sec.~\ref{sec:setup}, we describe the LBC data set and background models. In Sec.~\ref{sec:srdm_signal}, we describe out calculation of the SRDM signal. The signal calculation includes a dedicated treatment of dielectric screening in silicon, which becomes relevant in the boosted kinematic regime probed by SRDM. In Sec.~\ref{sec:statistical_analysis}, we present the statistical framework which we use to set constraints. Finally in Sec.~\ref{sec: discussion}, we present and discuss the resulting limits on the DM-electron scattering cross section. We consider dark-photon-mediated interactions and present results for the heavy- and ultralight-mediator scenarios.  




\section{Setup and data} 
\label{sec:setup}
The present analysis uses the same data sample as Ref.~\cite{DAMIC-M:2025}, differing only in the signal interpretation. We summarize the detector, operating conditions, and event selection, and then describe the background model used for the SRDM search.

The data were acquired with the DAMIC-M LBC, a prototype detector operated at the Modane Underground Laboratory~\cite{LBC:2024}. The LBC hosts two DAMIC-M prototype CCD modules implementing the final detector design. Each module contains four high-resistivity ($>10$~\unit{\kohm\cm}) n-type silicon skipper CCDs. Each CCD has an active area of $6144\times1536$ pixels, a pixel size of $15\times15$~\unit{\micrasq}, a thickness of $670$\unit{\micra}, and a mass of approximately $3.3$~\unit{\gram}. The CCDs are mounted inside a high-purity copper enclosure, which also shields them from infrared radiation. They are operated in a vacuum cryostat at a temperature of about $130$~\unit{\kelvin}. The cryostat is surrounded by lead and polyethylene shielding to reduce environmental backgrounds.

The CCDs are operated fully depleted at a substrate bias voltage of $45\,\unit{\volt}$. Charge carriers produced in the silicon bulk drift toward the pixel array under the applied electric field and undergo thermal diffusion during transport, producing a lateral spread that depends on the interaction depth. The CCD readout configuration used throughout this analysis corresponds to a $100\times1$ binning, in which 100 physical pixels are combined along the vertical direction into one recorded pixel. In the following, pixels refer to these recorded binned pixels. The pixel charge is read out with skipper amplifiers using multiple non-destructive measurements of the same charge packet, yielding a pixel-charge resolution $\sigma_{\rm pix}$ of approximately $0.16$\electron{}. Since images are read out without dead time, the exposure accumulated by each pixel is approximately equal to the image readout time, $1668$~\unit{\sec}.

The data set comprises about $84$~days of operation between October 2024 and January 2025. The first $\sim 7$~days of data (D1) were used to develop and validate the masking and selection procedures, while the remaining $\sim 77$~days (D2) were kept blind and used for the search. Two CCDs in Module 2 were excluded because of readout-related anomalies, leaving six CCDs in the final data sample. After all data-quality and masking requirements, $95\%$ of the active area is retained, corresponding to integrated exposures of $0.139$~kg-day and $1.257$~kg-day, for D1 and D2.

\subsection{Pattern Selection}
The event reconstruction and low-energy selections are identical to those used in Ref.~\cite{DAMIC-M:2025}, and are summarized here. In that procedure, consecutive images from each CCD are combined and calibrated using the fitted positions of the $0$\electron{} and $1$\electron{} peaks in the pixel-charge distribution. The masking and low-energy reconstruction steps remove defective structures, correlated noise, charge trailing, and localized regions with anomalous occupancies. Contiguous charged pixels are grouped into clusters, and clusters with a total reconstructed charge $\geq 6$\electron{} are excluded, since their expected contribution to the SRDM sensitivity in the search region is negligible.

In principle, a search for low-energy DM--electron interactions could be based directly on isolated single-pixel clusters. In this case, the background contribution to pixels with charge above $1$\electron{} could be estimated from the measured single-electron rate under a Poisson model of uncorrelated pixel charges. In the D1 control sample, however, the isolated $2$\electron{} and $3$\electron{} populations are not described by a single Poisson process normalized to the isolated $1$\electron{} rate. The origin of this non-Poissonian single-pixel population is still under investigation, but studies performed so far point to a substantial contribution from charge produced in the serial register, correlated with the readout clocking and voltage settings. To avoid relying on a background model for isolated pixels, the selection was fixed to exclude single-pixel clusters before unblinding the D2 sample.

The search is instead based on multi-pixel charge patterns, defined as clusters composed of two or three neighboring charged pixels. Ionization deposits occurring in the silicon bulk can produce charge shared across adjacent pixels because of diffusion during drift, whereas accidental coincidences of uncorrelated low-energy charges populate the same pattern channels with rates determined from the measured isolated-pixel occupancies in the $1$--$3$\electron{} range. This requirement discards any signal component reconstructed as isolated pixels, with the corresponding loss of acceptance included in the signal model, while retaining sensitivity to bulk ionization deposits through the multi-pixel pattern channels. For deposits in the $2$--$4$~\electron{} range, the corresponding pattern-selection acceptance is approximately 40\%--80\%, depending on the charge multiplicity, using the same reconstruction procedure as in Ref.~\cite{DAMIC-M:2025}.

Because of the $100\times1$ vertical binning used in the readout, only patterns extending along a row are considered. Selected events are classified into six charge-pattern channels:
\begin{equation}
    p \in \{11,\,21,\,111,\,31,\,22,\,211\},
\end{equation}
where each digit denotes the charge, in units of \electron{}, assigned to one pixel in the pattern, and channels with unequal digits include all permutations. The assignment of a given observed cluster to one of these channels is performed using the same pattern-identification procedure as in Ref.~\cite{DAMIC-M:2025}, which evaluates the compatibility of the observed pixel values with the corresponding integer-charge hypothesis, taking into account the measured pixel-charge resolution $\sigma_{\rm pix}$.

The observed counts in these six pattern channels constitute the search observable. In the unblinded D2 sample, the observed counts are
\begin{equation}
    D_p = (144,\,0,\,0,\,1,\,0,\,0),
\end{equation}
for $p=(11,\,21,\,111,\,31,\,22,\,211)$. For the SRDM reinterpretation, the selected patterns are organized by image and pattern channel, with $D_{p,i}$ denoting the observed count in channel $p$ for image $i$. The exposure-integrated counts quoted above correspond to $D_{p}=\sum_{i}D_{p,i}$, and quantities without an image index follow this convention. This image-level representation is used throughout the analysis to construct the expected signal and background yields either image by image or after summing over the blind D2 exposure.

\subsection{Background Model}
The nominal background prediction is constructed image by image. For image $i$ and pattern channel $p$, it is written as the sum of two components,
\begin{equation}
    B_{p,i}^{\mathrm{nom}} = B_{p,i}^{\mathrm{rc}} + B_{p,i}^{\mathrm{rad}},
\end{equation}
where $B_{p,i}^{\mathrm{rc}}$ denotes the contribution from random coincidences of uncorrelated low-charge pixels, and $B_{p,i}^{\mathrm{rad}}$ is the contribution from radiogenic interactions in the detector and surrounding apparatus. As for the observed counts, quantities without an image index denote sums over the blind D2 exposure. The background model is constructed and validated using the D1 control sample, and all selection and background-modeling choices are fixed before unblinding the D2 search sample.\\

The random-coincidence component is derived from isolated-pixel rates measured in the D1 control sample in consecutive 10-column groups, to account for column-dependent non-uniformities across the sensors. For each image $i$, these rates define the Poisson probability $P_{q,i}^{(c)}$ for an accidental low-charge configuration $q$ to occur in column group $c$ during the image exposure. The expected number of random coincidences reconstructed in pattern channel $p$ for image $i$ is
\begin{equation}
    B_{p,i}^{\mathrm{rc}} =
    \sum_{c}\sum_{q\in\mathcal{Q}_p}
    N_{q,c}^{i}\, P_{q,i}^{(c)}\, \epsilon_{q\rightarrow p},
\end{equation}
where $N_{q,c}^{i}$ is the effective number of available adjacent-pixel positions for configuration $q$ in column group $c$, after applying the image mask. The set $\mathcal{Q}_p$ includes both the nominal configuration, $q=p$, and alternative low-charge configurations that can be reconstructed as channel $p$ because of the finite pixel-charge resolution $\sigma_{\mathrm{pix}}$. The corresponding probabilities $\epsilon_{q\rightarrow p}$ that configuration $q$ is reconstructed and identified as channel $p$ are evaluated with toy Monte Carlo simualtions. 
Using the same pattern ordering as for $D_p$, the random-coincidence expectations summed over the blind D2 exposure are
\begin{equation}
    B_{p}^{\mathrm{rc}} =
    (141.4,\,0.111,\,0.042,\,0.019,\,2.5\times10^{-5},\,5.8\times10^{-5}).
\end{equation}\\

The second component accounts for physical ionization deposits from radiogenic and external background sources. This contribution is included because the external lead-and-polyethylene shielding was not operated in its nominal fully closed configuration throughout the science exposure, leaving a residual contribution from ambient $\gamma$-rays and neutrons.

The nominal normalization of this component is estimated from the differential cluster rate measured in a higher-energy control region, $R_{\mathrm{ctrl}}$, where random coincidences are negligible and the sample is dominated by physical backgrounds. The Monte Carlo--simulated $\gamma$-background model developed for the LBC setup~\cite{LBC:2024} is used to determine the transfer factor $\tau$ between this control-region rate and the differential rate of radiogenic deposits in the low-energy DM search region. From the simulations, this factor is $\tau\simeq0.15$.
The expected number of radiogenic deposits in each integer-charge bin is obtained using the mean electron-hole creation energy in silicon at the working temperature, $\epsilon_{\rm eh}=3.72$~eV, and the effective D2 exposure of each image. These deposits are folded with the diffusion and pattern-selection probabilities $\epsilon_{n_e\rightarrow p}$ for a deposit of total charge $n_e=2,\ldots,5$~\electron{} to be reconstructed in pattern channel $p$. 
The expected radiogenic contribution in pattern channel $p$ and image $i$ is therefore
\begin{equation}
    B_{p,i}^{\mathrm{rad}}
    =
    R_{\mathrm{ctrl}}\,\tau\,\mathcal{E}^{i}_{\mathrm{D2}}\,\epsilon_{\rm eh}
    \sum_{n_e=2}^{5}\epsilon_{n_e\rightarrow p}.
\end{equation}
where $R_{\mathrm{ctrl}}$ is the differential rate measured in the control region. 

Using the same pattern ordering as for $D_p$, the resulting radiogenic expectations summed over the blind D2 exposure are
\begin{equation}
    B_{p}^{\mathrm{rad}} =
    (0.039,\,0.039,\,0.016,\,0.052,\,0.011,\,0.035).
\end{equation}

Combining the two components, the background expectation for image $i$ and pattern channel $p$ is parameterized as
\begin{equation}
    B_{p,i}(\theta_{\mathrm{rad}})
    =
    B_{p,i}^{\mathrm{rc}}
    +
    \Theta_{\mathrm{rad}} B_{p,i}^{\mathrm{rad}},
    \label{eq:bkg_model}
\end{equation}
where $B_{p,i}^{\mathrm{rad}}$ is the nominal radiogenic expectation obtained from the control-region normalization, corresponding to $\Theta_{rad}=1$. The nuisance parameter $\Theta_{\mathrm{rad}}$ rescales the radiogenic component and is constrained by the observed control-region population through an auxiliary Poisson term in the likelihood, as described in Sec.~\ref{sec:statistical_analysis}.
\section{Solar Reflected Dark Matter Signal }
\label{sec:srdm_signal}

\subsection{Solar Reflected Flux}
\label{sec:srdm_flux}
The SRDM flux at Earth was generated with the \damascussun{} Monte Carlo code~\cite{DaMaSCUS-SUN:2022}, using its dark-photon implementation for sub-GeV dark matter~\cite{DaMaSCUS-SUN:2024}. Flux simulations were performed for the heavy- and ultralight-mediator benchmarks on a discrete grid in $(m_{\chi}, \bar{\sigma}_e)$. For each point in this grid, approximately $N_{\rm sim} \sim 10^{6}$ incident Galactic-halo particles were propagated through the Sun. The \damascussun{} propagation accounts for the Sun's motion through the Galactic halo, gravitational focusing, and stochastic scatterings of halo DM particles with thermal solar electrons and nuclei, modeled with local Maxwell-Boltzmann velocity distributions. The incident halo population is described using the SHM parameters recommended in Ref.~\cite{Baxter:2021pqo}. For dark-photon-mediated interactions, the solar plasma modifies the effective kinetic mixing through in-medium polarization effects, which we include in the simulations.

The solar-reflected population is defined by particles that scatter at least once in the Sun and subsequently escape the Sun. This population is used to construct the speed-differential flux at Earth's orbit,
\begin{equation}
\frac{d\Phi_\chi}{dv} = \frac{1}{4\pi \ell^2}\frac{N_{\rm refl}} {N_{\rm sim}}\, \Gamma_\odot \, f_{\rm refl}(v).
\label{eq_srdm_flux}
\end{equation}
Here $\ell$ is the Sun--Earth distance, $N_{\rm refl}/N_{\rm sim}$ is the reflected fraction for the simulated configuration, $\Gamma_{\odot}$ is the rate of halo DM particles incident on the Sun, and $f_{\rm refl}(v)$ is the normalized speed distribution of the reflected population.

The speed-differential flux in Eq.~\eqref{eq_srdm_flux} is parametrized by the isoreflection angle $\theta_{\rm iso}$, which accounts for the anisotropy of the solar-reflected flux. This angle is defined by the orientation of the Earth--Sun direction with respect to the Sun's motion through the Galactic halo. This angle varies over the course of a year, as the Earth orbits the Sun. Our LBC exposure covers a range of isoreflection angles of approximately $\theta_\mathrm{iso} = [60^\circ,\,120^\circ]$. For each $(m_{\chi}, \bar{\sigma}_e)$ grid point, the flux was generated for \num{18} discrete values of $\theta_{\rm iso}$, referred to here as angular rings.  The UTC timestamp of each image determines the corresponding value of $\theta_{\rm iso}$, allowing an image-dependent SRDM flux prediction to be constructed by interpolation between neighboring rings. The angular interpolation was validated by comparing simulations with different numbers of rings. 

\begin{figure}[!hbt]
    \centering
    \includegraphics[width=\linewidth]{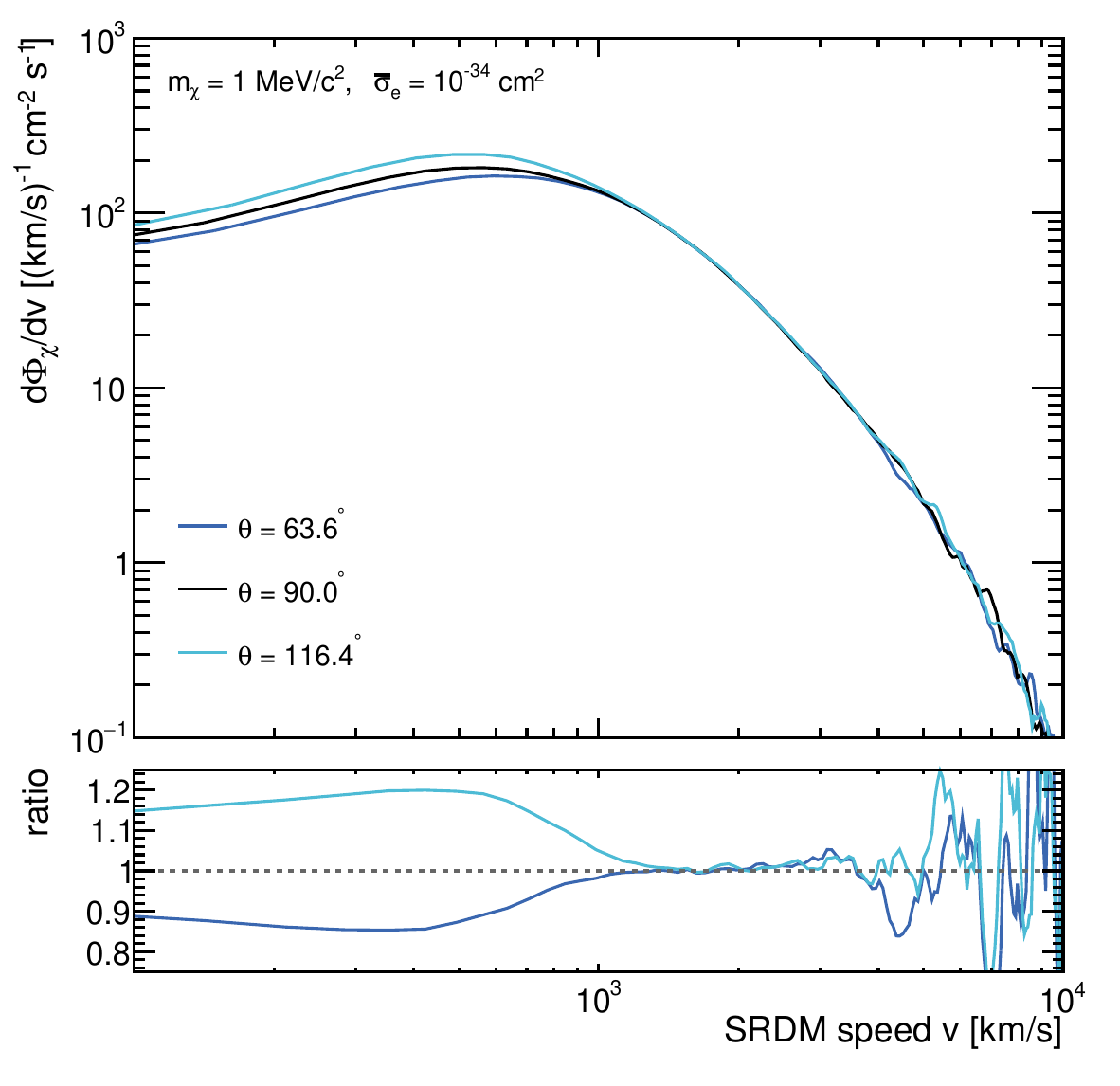}
    \caption{Dependence of the SRDM differential flux on the isoreflection angle $\theta_{\rm iso}$ for a 
    representative $(m_{\chi}, \bar{\sigma}_e)$ grid point. The upper panel shows the spectra for three angular 
    rings relevant to the LBC exposure, and the lower panel shows their ratio to a reference spectrum. 
    For this representative point, the spectra show only a mild dependence on $\theta_{\rm iso}$ over the angular range sampled by the LBC exposure. Fluctuations in the high-velocity tail are dominated by finite Monte Carlo statistics.}
    \label{fig:srdm_flux_theta_variation}
\end{figure}

The dependence of the flux on the isoreflection angle is illustrated in Fig.~\ref{fig:srdm_flux_theta_variation} for a representative parameter-space point. 
To quantify the impact of the annual variation, the limit calculation was performed with both the image-dependent signal prediction and with a time-independent SRDM flux, evaluated at $\theta_{\rm iso}=86.9^{\circ}$, the mean isoreflection angle of the dataset. As discussed in Sec.~\ref{sec:statistical_analysis}, the difference between the image-dependent and time-independent treatments is negligible for both mediator benchmarks over the LBC exposure. We therefore assume a fixed value of $\theta_{\rm iso}^{\rm ref}=86.9^{\circ}$ for our nominal signal model. 

\begin{figure}[!htb]
    \centering
    \includegraphics[width=\linewidth]{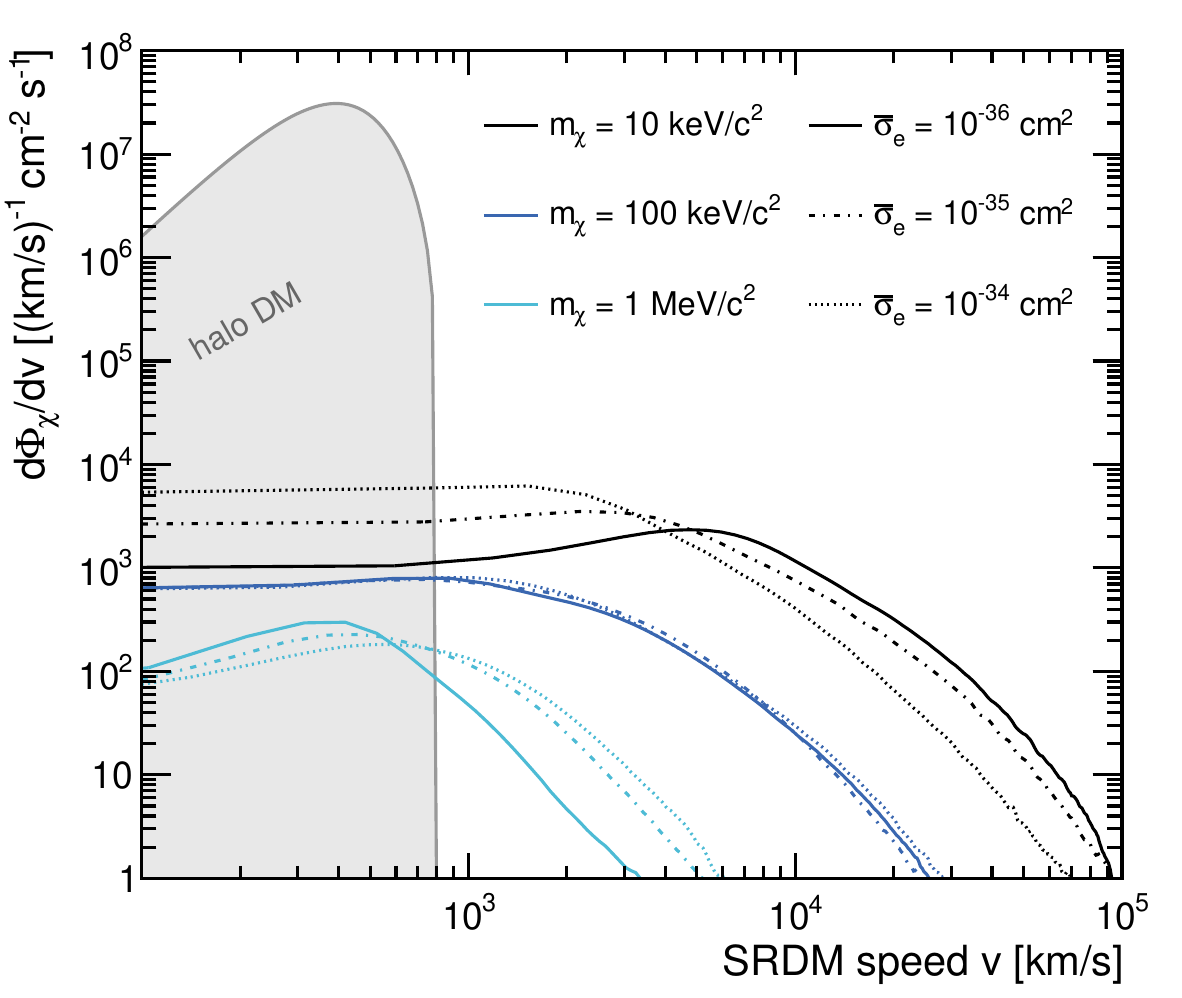}
    \caption{SRDM differential flux at Earth's orbit for the ultralight-mediator benchmark, shown for selected DM masses and reference cross sections at $\theta_{\rm iso}^{\rm ref}$. The shaded region indicates the characteristic support of the standard halo speed distribution. The spectra extend beyond the standard halo velocity range and vary in both normalization and shape with $\bar{\sigma}_{e}$, reflecting the nonlinear dependence of the escaping reflected flux on the scattering cross section.}
    \label{fig:srdm_flux_velocity}
\end{figure}

Representative SRDM speed distributions are shown in Fig.~\ref{fig:srdm_flux_velocity} for selected masses and DM-electron cross sections. The shaded region indicates the characteristic flux of the standard Galactic halo distribution. Solar reflection produces a non-thermal, parameter-dependent flux extending well beyond the halo velocity range, thereby opening otherwise inaccessible phase space for direct detection. The figure also illustrates that changing $\bar{\sigma}_e$ modifies both the normalization and the spectral shape of the reflected component. Varying the cross-section changes the opacity of the Sun to DM, affecting whether the DM scatters predominantly in the hotter core or in the cooler atmosphere, and subsequently the maximum velocity which the SRDM can attain. 
The reflected flux is therefore tabulated over the simulated ($m_{\chi}$,$\bar{\sigma}_{e}$) grid and interpolated during the limit calculation, rather than obtained by rescaling a fixed reference spectrum.

Additional propagation of the SRDM flux through the Earth is not included in the nominal rate calculation. 
This effect would introduce a diurnal dependence through the solar zenith angle at the detector and could attenuate or distort the already-reflected flux along its path to the detector. Dedicated checks with \textsc{Verne2}~\cite{Verne2} show that Earth-propagation effects produce a negligible change in the expected signal yield in the ultralight-mediator parameter space driving the sensitivity of this analysis. Larger variations can occur in parts of the heavy-mediator parameter space. However, the regions where this effect is most relevant are already strongly constrained by existing searches, and including a fully time-dependent Earth-propagation correction would not change the physics conclusions of the present work. The nominal rate calculation therefore uses the SRDM flux at Earth's position, without propagation through the terrestrial overburden.

\subsection{DM-electron scattering rate in silicon}
\label{sec:dme_rate}

The SRDM flux defined in Sec.~\ref{sec:srdm_flux} is converted into the DM--electron scattering rate in silicon using a modified \qcdark{} calculation~\cite{Dreyer:2023ovn}. 
For an electronic excitation of energy $E_e$ produced with momentum transfer $q$, the relevant dependence of the semiconductor scattering rate can be written schematically as~\cite{Essig:2015cda}:
\begin{equation}
\begin{aligned}
    \left.
    \frac{dR}{dE_e}(m_\chi,\bar{\sigma}_e)
    \right|_{\theta_{\rm iso}=\theta_{\rm iso}^{\rm ref}}
        &\propto
        \int \frac{dq}{q^2}\,
        \eta\!\left(v_{\min}\right)
        \left|F_{\rm DM}(q)\right|^2  \\
        &\quad\times
        \left|F_c(q,E_e)\right|^2
        \left|\epsilon(q,E_e)^{-1}\right|^2 .
\end{aligned}
\label{eq:dR_dEe}
\end{equation}
Here $\theta_{\rm iso}^{\rm ref}$ is the reference value defined in Sec.~\ref{sec:srdm_flux}. The factor $F_{\rm DM}(q)$ describes the mediator dependence; $F_c(q,E_e)$ is the silicon crystal form factor computed by \qcdark{}; and $\epsilon(q,E_e)$ is the dielectric response of the silicon target. The implementation of this screening factor for the SRDM rate calculation is discussed in the next section. The $\eta$ factor is the SRDM analogue of the usual mean inverse speed~\cite{Cerdeno:2010jj}, or velocity integral, evaluated from the SRDM flux rather than from a standard halo velocity distribution.

The interaction is assumed to be mediated by a dark photon of mass $m_{A'}$. In the heavy-mediator limit, $m_{A'} \gg q$, the interaction is effectively contact-like and $F_{\rm DM}(q)=1$. In the ultralight-mediator limit, $m_{A'} \ll q$, the propagator gives $F_{\rm DM}(q) = (\alpha m_e / q)^{2}$, where $m_e$ is the electron mass and $\alpha$ is the fine-structure constant.

The inverse-speed integral $\eta$ contains the kinematic dependence of the incident SRDM population above the minimum speed $v_{\min}(q,E_e)$ required to produce an excitation of energy $E_e$ with momentum transfer $q$. For the nominal calculation, it is evaluated using the SRDM flux defined at $\theta_{\rm iso}^{\rm ref}$:
\begin{equation}
    \eta \left(v_{\min}; m_\chi, \bar{\sigma}_e, \theta_{\rm iso}^{\rm ref} \right)
            =
            \int_{v_{\min}}^{v_{\rm max}^{\rm SRDM}}
            \frac{dv}{v^{2}}\,
            \left. 
            \frac{d\Phi_\chi}{dv}
            \right|_{\theta_{\rm iso}=\theta_{\rm iso}^{\rm ref}}
            \,.
    \label{eq:srdm_eta}
\end{equation}
The image-dependent variant is obtained by replacing $\theta_{\rm iso}^{\rm ref}$ with the value $\theta_{\rm iso}(t_i)$ corresponding to each image.

\subsection{Dielectric screening in the SRDM regime}
\label{sec:screening}
The dielectric factor $|\epsilon^{-1}(q,E_e)|^{2}$ in Eq.~\eqref{eq:dR_dEe} describes the in-medium response of the silicon target to the dark-photon-mediated interaction. This correction enters the detector-rate calculation and is separate from the in-medium dark-photon mixing included in the solar propagation, which modifies the scattering probabilities in the solar plasma and therefore the reflected flux. For the boosted SRDM flux considered here, the dielectric response must be evaluated over a broader region of the $(q,E_e)$ plane than in standard halo calculations, including finite excitation energies and relatively low momentum transfers. The default \qcdark{} implementation~\cite{Essig:2015cda} uses a Thomas--Fermi-like screening prescription. This approximation captures static low-momentum screening, but does not describe the finite-frequency dielectric response of silicon. 

A finite-frequency dielectric model is therefore required. The Lindhard dielectric function captures the momentum- and energy-dependent response of a homogeneous electron gas~\cite{lindhard1954}, but without dissipation it produces an artificially narrow plasmon feature, as it can be seen in Fig \ref{fig:screening}. We therefore use the Mermin prescription, which extends the Lindhard response by introducing a finite damping parameter while preserving charge conservation~\cite{Mermin:1970zz}. The damping parameter broadens the plasmon peak and provides a more physical description of the dissipative response of silicon. We adopt $\Gamma=3.2\,{\rm eV}$, corresponding to the measured width of the silicon plasmon peak around $E_p\simeq16.6\,{\rm eV}$~\cite{eels}. 

The explicit expressions used for the Mermin and Lindhard dielectric functions, together with the numerical conventions used in the implementation, are given in Appendix~\ref{app:screening_formalism}. The nominal SRDM signal calculation evaluates the dielectric factor in Eq.~\eqref{eq:dR_dEe} using the Mermin response, $\epsilon_{\rm Mer}(q,E_e)$ defined in Eq.~\eqref{eq:app_nominal_screening}, in place of the default \qcdark{} screening prescription.

\begin{figure}[tb]
    \centering
    \includegraphics[width=\linewidth]{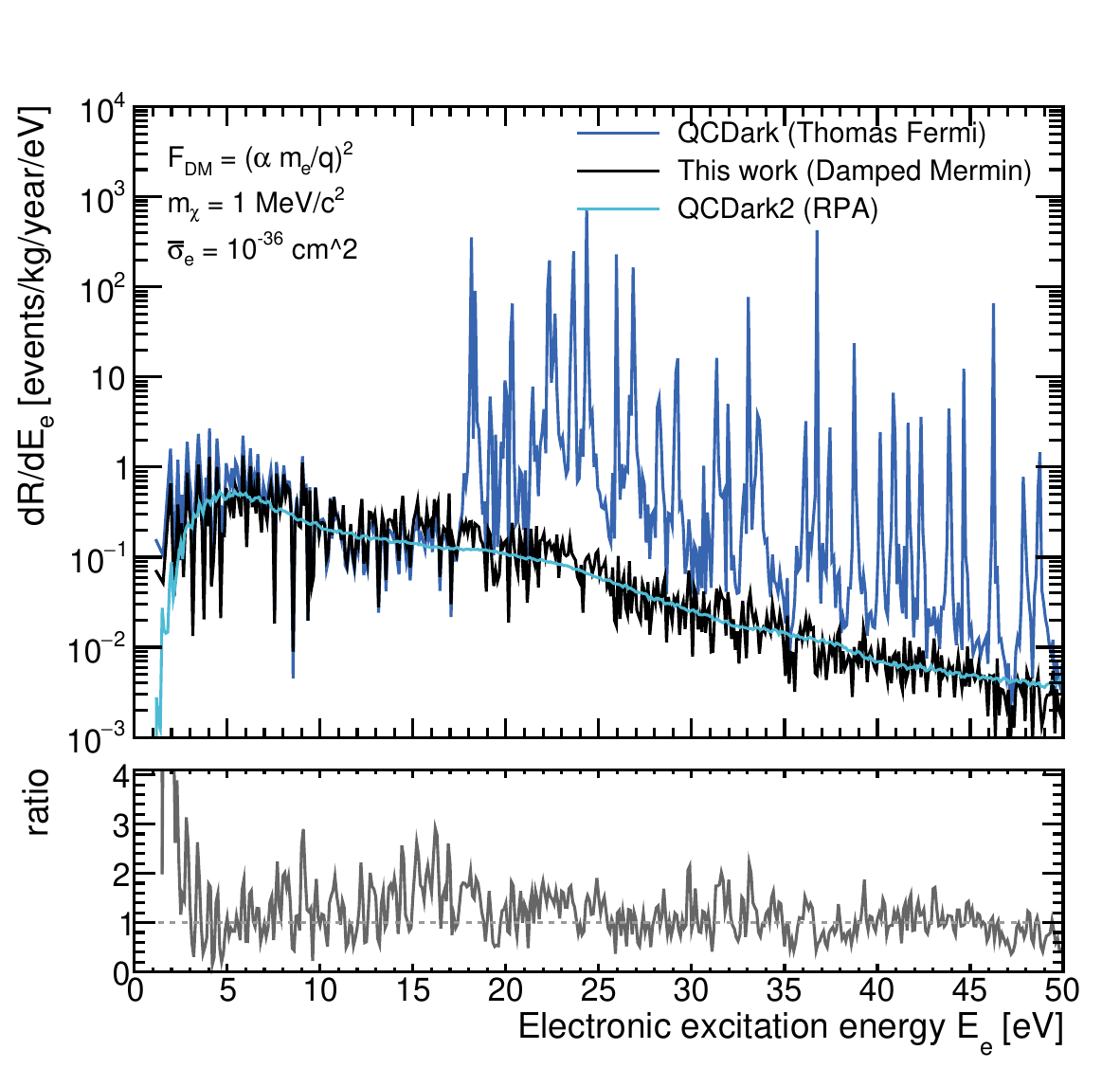}
    \caption{Impact of the dielectric-screening prescription on the predicted SRDM electronic excitation rate in silicon for the ultralight-mediator benchmark indicated in the figure. The original \qcdark{} calculation with Thomas--Fermi screening is compared with the nominal calculation used in this work, which retains the \qcdark{} crystal form factor but replaces the screening factor by the Mermin dielectric response, and with the recent \qcdarktwo{} prediction based on an RPA dielectric response. The lower panel shows the ratio of the nominal calculation to \qcdarktwo{}.}
    \label{fig:screening_rate_comparison}
\end{figure}

The impact of the screening prescription on the predicted SRDM rate is shown in Fig.~\ref{fig:screening_rate_comparison}. The comparison includes the original \qcdark{} calculation using its default Thomas--Fermi screening prescription, the nominal calculation used in this work based on the Mermin dielectric response, and the recent \qcdarktwo{} prediction~\cite{Dreyer:2026bmz} based on an RPA dielectric response. The static prescription produces pronounced structures in the SRDM regime, associated with the plasmon-region contribution. Replacing it with the damped Mermin response regularizes this contribution and yields a rate consistent with the \qcdarktwo{} prediction in the energy range relevant for this analysis. The residual bin-to-bin fluctuations visible in the nominal curve are numerical artifacts associated with the finite resolution of the crystal form-factor tables and phase-space integration inherited from the original \qcdark{} implementation, rather than physical structures in the dielectric response. The results from \qcdarktwo{} and the damped Mermin model used are shown to be compatible, so it makes for a valid alternative before the full migration to \qcdarktwo{} as the standard software for DM-electron scattering calculations.

\subsection{SRDM Signal in Pattern Space}
\label{section:srdm_pattern_rates}

The final step in the signal model is to convert the differential SRDM rate in silicon, defined in Sec.~\ref{sec:dme_rate}, into an expected signal yield in the charge-pattern channels defined in Sec.~\ref{sec:setup}. This conversion accounts for the ionization-yield model, charge diffusion in the CCD, the 100×1 binning, and the pattern-identification efficiency.

The deposited energy $E_e$ is converted into a discrete number of electron--hole pairs, $n_{eh}$, using the probability distribution $\mathcal{P}(n_{eh}|E_e)$ from the semi-empirical model of Ref.~\cite{Ramanathan:2020}. The detector response is encoded in $\mathcal{P}(p|n_{eh})$, the probability that an interaction producing $n_{eh}$ electron--hole pairs is reconstructed in pattern channel $p$ by the pattern-identification procedure described in Ref.~\cite{DAMIC-M:2025}. This probability is obtained from Monte Carlo simulations of charge diffusion in the CCD, including depth-dependent lateral diffusion, the $100\times1$ binning used during data taking, and the measured pixel readout noise $\sigma_{\rm pix}$.

The expected signal contribution in pattern channel $p$ and image $i$ is then
\begin{equation}
\begin{aligned}
    S_{p,i}(m_{\chi},\bar{\sigma}_e)
    & =
    \mathcal{E}^{i}_{\mathrm{D2}}
    \sum_{n_{eh}}
    \mathcal{P}(p|n_{eh})\\
    &\quad \times
    \int dE_e\,
    \left.
    \frac{dR}{dE_e}
    \right|_{\theta_{\rm iso}=\theta_{\rm iso}^{\rm ref}}
    \mathcal{P}(n_{eh}|E_e),
    \label{eq:signal_pattern_template}
\end{aligned}
\end{equation}
where $\mathcal{E}^{i}_{\mathrm{D2}}$ is the effective exposure of image $i$ after data-quality and masking selections. In the nominal signal model, the differential rate is common to all images for fixed $(m_\chi,\bar{\sigma}_e)$; the image index in $S_{p,i}$ accounts for the image-dependent exposure. The corresponding time-dependent signal prediction is obtained by replacing the reference-angle rate with the rate evaluated at the value of $\theta_{\rm iso}$ associated with each image, as described in Sec.~\ref{sec:srdm_flux}.

\begin{figure}[tb]
    \centering
    \includegraphics[width=\linewidth]{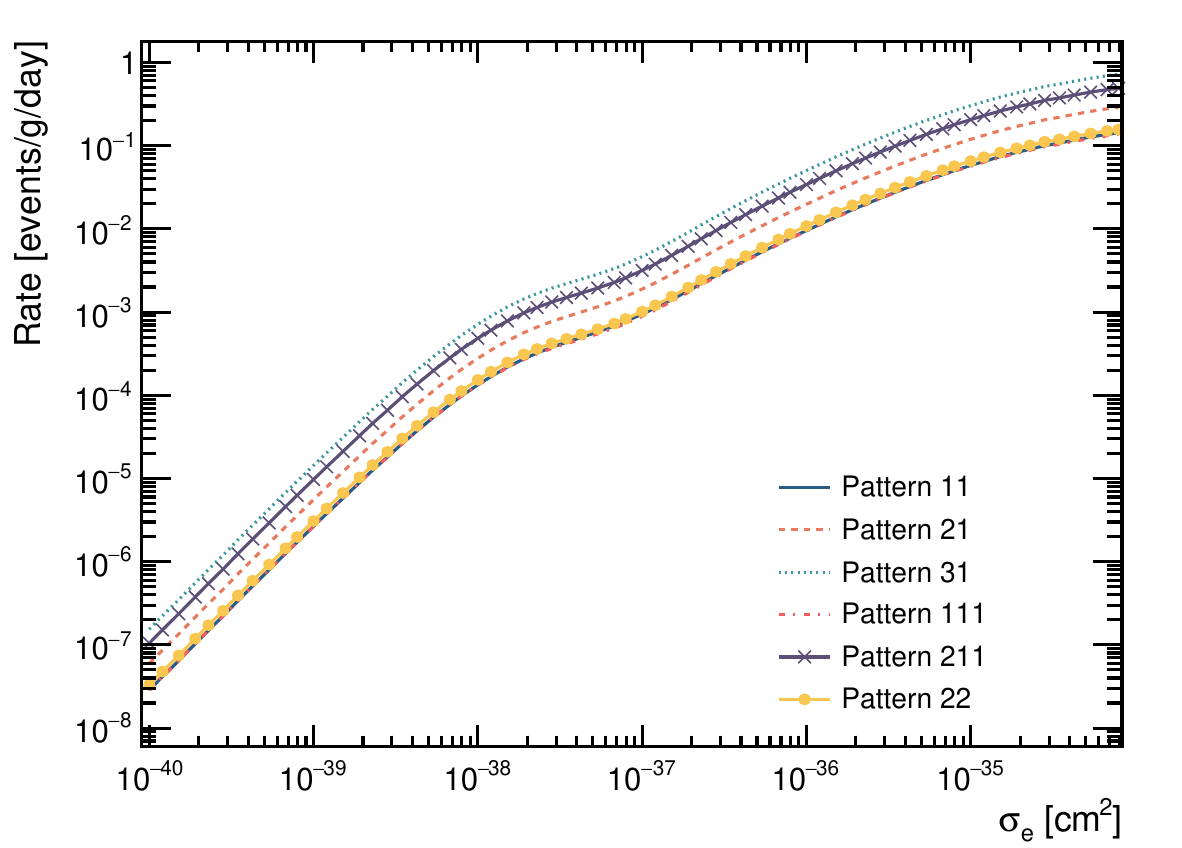}
    \caption{Expected SRDM signal rate in events/g/day for $m_{\chi}=0.01$~\MeV{} as a function of the reference DM--electron scattering cross section, evaluated with the nominal signal model. All six pattern channels used in the analysis are shown.}
    \label{fig:signal_vs_crosssection}
\end{figure}

Combining the signal and background contributions, the expected number of events for image $i$ and pattern channel $p$ is
\begin{equation}
    \nu_{p,i}(m_{\chi},\bar{\sigma}_e,\Theta_{\rm rad})
    =
    S_{p,i}(m_{\chi},\bar{\sigma}_e)
    +
    B_{p,i}(\Theta_{\rm rad}).
    \label{eq:expected_events}
\end{equation}

\section{Statistical analysis}
\label{sec:statistical_analysis}
For each fixed DM mass and mediator benchmark, the likelihood is evaluated as a function of the tested reference DM--electron scattering cross section, denoted by $\mu\equiv\bar{\sigma}_e$. 
The statistical inference is based on the six charge-pattern counts defined in Sec.~\ref{sec:setup}. The SRDM signal prediction is evaluated at the reference isoreflection angle $\theta_{\rm iso}^{\rm ref}$ and integrated over the blind D2 exposure. The expected number of events in pattern channel $p$, denoted $\nu_p^{\rm ref}(\mu,\Theta_{\rm rad})$, is obtained from Eq.~\eqref{eq:expected_events} by summing over images, with the background model of Eq.~\eqref{eq:bkg_model}.

The likelihood for the six search charge-pattern channels is
\begin{equation}
    \mathcal{L}_{\rm search}(\mu,\Theta_{\rm rad}) = \prod_p {\rm Pois}
    \left( D_p\,\middle|\, \nu_p^{\rm ref}(\mu,\Theta_{\rm rad}) \right).
    \label{eq:likelihood_search_ref}
\end{equation}
The radiogenic-background normalization is constrained by the higher-energy control region introduced in Sec.~\ref{sec:setup}. This constraint is implemented through the auxiliary Poisson term
\begin{equation}
    \mathcal{L}_{\rm ctrl}(\Theta_{\rm rad}) = {\rm Pois}
        \left(N_{\rm ctrl}^{\rm obs}\,\middle|\, \Theta_{\rm rad} N_{\rm ctrl}^{\rm nom}\right),
    \label{eq:likelihood_ctrl}
\end{equation}
where $N_{\rm ctrl}^{\rm obs}=98$ is the observed control-region population and $N_{\rm ctrl}^{\rm nom}$ is the nominal control-region expectation corresponding to $\Theta_{\rm rad}=1$. The full likelihood used for the reference treatment is
\begin{equation}
    \mathcal{L}(\mu,\Theta_{\rm rad}) =
    \mathcal{L}_{\rm search}(\mu,\Theta_{\rm rad}) \mathcal{L}_{\rm ctrl}(\Theta_{\rm rad}) \,.
    \label{eq:likelihood_full}
\end{equation}

To test each value of $\mu$ and construct upper limits in the absence of a significant signal preference, we follow the profile-likelihood approach of Ref.~\cite{Cowan:2013}. The one-sided test statistic is defined as
\begin{equation}
    \tilde q_{\mu} =
    \begin{cases}
    -2\ln \dfrac{\mathcal{L}(\mu,\widehat{\widehat{\Theta}}_{\mu})}
            {\mathcal{L}(0,\widehat{\widehat{\Theta}}_{0})},
    & \widehat{\mu}<0,
    \\[0.35cm]
    -2\ln \dfrac{\mathcal{L}(\mu,\widehat{\widehat{\Theta}}_{\mu})}
            {\mathcal{L}(\widehat{\mu},\widehat{\Theta})},
    & 0\leq\widehat{\mu}\leq\mu,
    \\[0.35cm]
    0,
    & \widehat{\mu}>\mu .
    \end{cases}
    \label{eq:test_statistic_onesided}
\end{equation}
Here $\widehat{\widehat{\Theta}}_{\mu}$ denotes the value of the nuisance parameter that maximizes the likelihood for fixed $\mu$, while $(\widehat{\mu},\widehat{\Theta})$ denotes the global maximum-likelihood estimate. The first branch accounts for cases in which the unconstrained estimator of the cross section lies below the physical boundary at $\mu=0$, whereas the last branch makes the statistic insensitive to upward fluctuations beyond the tested cross section.

The \(p\)-value for a given \(\mu\) is
\begin{equation}
    p_{\mu}
    =
    \int_{\tilde{q}_{\mu}^{\rm obs}}^{\infty}
    f(\tilde{q}_{\mu}\,|\,\mu)\,
    d\tilde{q}_{\mu},
    \label{eq:pvalue_twosided_v2}
\end{equation}
where \(f(\tilde{q}_{\mu}\,|\,\mu)\) is the sampling distribution of the test statistic under the signal-plus-background hypothesis with cross section \(\mu\).

Values of $\mu$ for which $p_{\mu}<0.10$ are excluded at $90\%$ confidence level, and the upper limit is obtained from $p_{\mu}=0.10$. Because of the low event counts in the pattern channels and the discreteness of the observable, asymptotic approximations are not used. The $p_{\mu}$ values are instead computed from toy Monte Carlo pseudo-experiments, as the fraction of experiments with $\tilde q_{\mu}\geq\tilde q_{\mu}^{\rm obs}$. For each tested $\mu$, pseudo-data are generated from Poisson distributions with means $\nu_p^{\rm ref}(\mu,\Theta_{\rm rad}=1)$. Each pseudo-experiment is then fitted with the same profiling procedure as the data.

The impact of the annual variation of the SRDM flux is evaluated with an otherwise identical statistical procedure. In this alternative treatment, the reference-angle signal prediction is replaced by the image-dependent prediction obtained by evaluating the SRDM flux at $\theta_{\rm iso}(t_i)$ for each image. The corresponding image-level likelihood is used only to quantify the effect of the time dependence. We define
\begin{equation}
    \alpha(m_{\chi}) = \frac{\mu_{90}^{\rm img}(m_{\chi})}{\mu_{90}^{\rm ref}(m_{\chi})},
    \label{eq:alpha_time_dependence}
\end{equation}
where $\mu_{90}^{\rm ref}$ is the upper limit obtained with the reference-angle signal prediction, and $\mu_{90}^{\rm img}$ is the corresponding result obtained with the image-dependent flux prediction. The ratio is found to be consistent with unity for both mediator benchmarks. The same conclusion is obtained for representative search-channel configurations spanning the range of upper limits covered by the observed result and the expected sensitivity bands, with differences remaining below the percent level in all cases considered. This validates the use of the reference-angle treatment in the final statistical interpretation.

\begin{figure*}[!htb]
    \centering
    \includegraphics[width=0.45\linewidth]{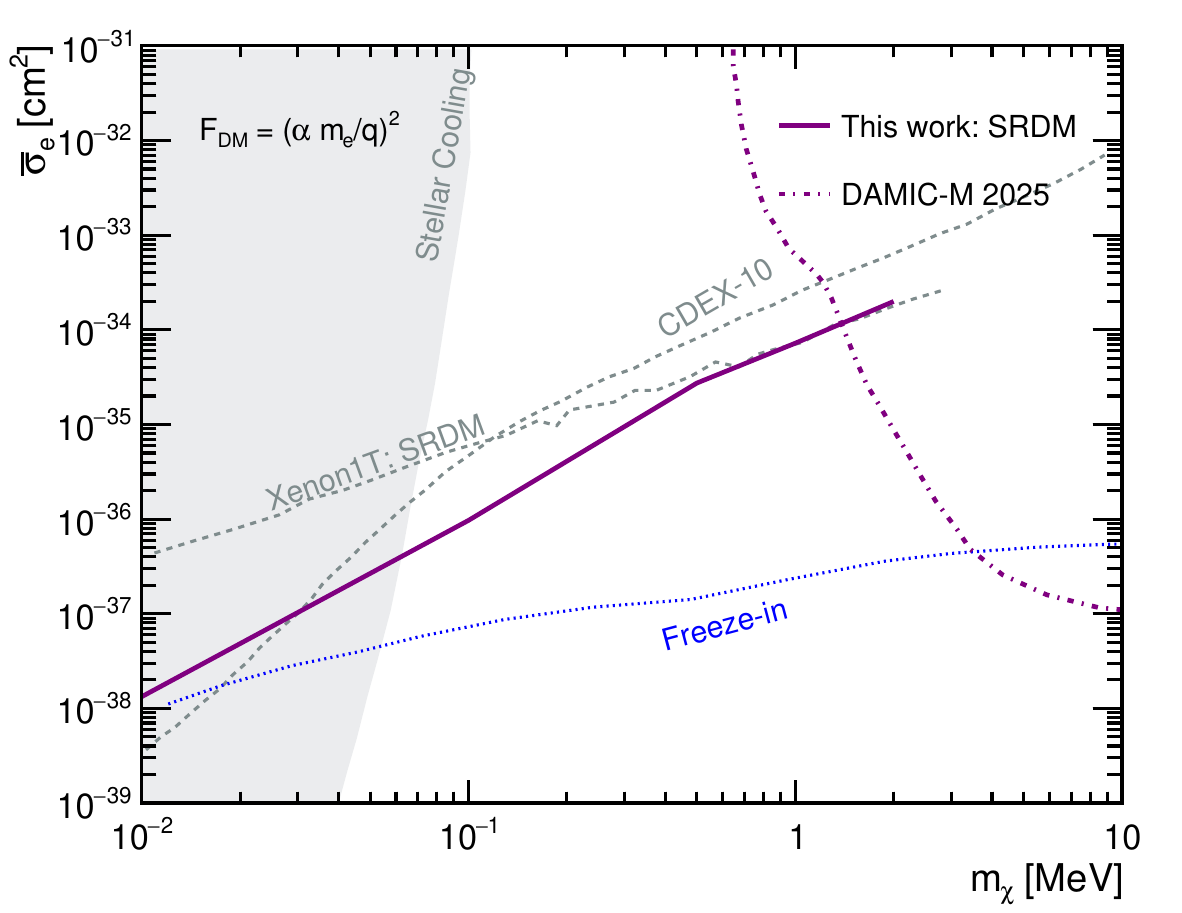}
    \hspace{0.6cm}
    \includegraphics[width=0.45\linewidth]{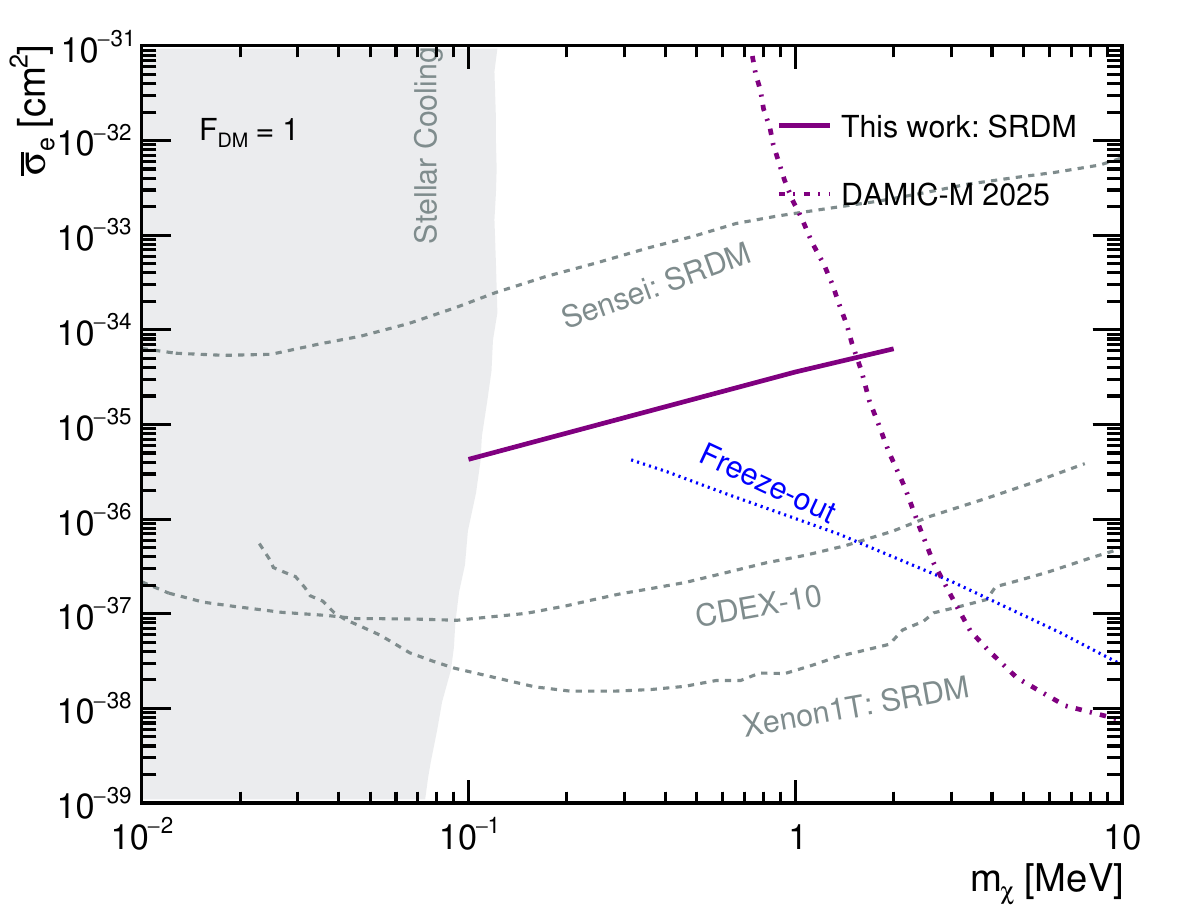}
    \caption{DAMIC-M 90\% C.L. upper limits on DM-electron interactions through an ultralight (left) and heavy (right) dark photon mediator obtained from the SRDM analysis. Also shown are previous DAMIC-M limits \cite{DAMIC-M:2025ltz, DAMIC-M:2025} merged into one limit grouping the best exclusion limits for DAMIC-M. Solar-reflected DM limits (such as CDEX-10 \cite{CDEX:2023wfz} and XENON1T \cite{xenoncollaboration2025constraintssolarreflecteddark}) are also shown in light gray dashed curves. The gray shaded region corresponds to cosmological constrains from stellar cooling ~\cite{stellar_constrains}. Theoretical expectations assuming a DM relic abundance from freeze-in and freeze-out mechanisms are shown in blue \cite{freezein}.} 
    \label{fig:upper_limits}
\end{figure*}

No statistically significant preference for an SRDM signal is found for either mediator benchmark. We therefore derive 90\% C.L. upper limits on the reference DM--electron scattering cross section, $\bar{\sigma}_e$, as a function of the DM mass. The resulting limits are shown in Fig.~\ref{fig:upper_limits}. These limits are weaker than the median expectation but lie just above the 1$\sigma$ sensitivity band over the mass range where the analysis is sensitive.

\section{Discussion}
\label{sec: discussion}

We have presented a dedicated search for solar-reflected dark matter using the DAMIC-M LBC dataset. The analysis exploits the sub-electron charge resolution of skipper CCDs and the charge-pattern search strategy developed for the halo-DM analysis, but targets a distinct signal population: dark-sector particles that scatter in the Sun and reach the detector with velocities above those expected from the standard Galactic halo.

This solar-reflected component provides a complementary probe of sub-MeV DM--electron interactions. The kinematic boost acquired in the Sun extends the direct-detection reach to the DM masses below the range accessible to conventional halo-DM electron-scattering searches, while testing parameter space complementary to astrophysical constraints from stellar cooling ~\cite{stellar_constrains}. In this sense, SRDM searches connect low-threshold direct-detection experiments to hidden-sector parameter space that is difficult to access with terrestrial halo-DM searches alone.

No statistically significant preference for an SRDM signal is observed in either ultralight- or heavy-mediator benchmark. We therefore set 90\% C.L. upper limits on the reference DM--electron scattering cross section, $\bar{\sigma}_{e}$. The signal prediction is computed with a modified implementation of \qcdark{}, including the dielectric-screening treatment adopted for the boosted-DM kinematic regime. Since the solar-reflected flux depends on $\bar{\sigma}_{e}$, the expected signal is evaluated independently at each tested cross section rather than obtained by rescaling a fixed signal template.

For the ultralight-mediator benchmark, the resulting limits provide the strongest sensitivity among the published direct-detection constraints shown in Fig.~\ref{fig:upper_limits} over the approximate mass range $0.05 - 0.5$~\MeV{}. 
For the heavy-mediator benchmark, liquid-noble experiments remain more sensitive in the boosted-DM parameter space, and the present results does not improve upon the leading constraints. 


The final DAMIC-M detector will provide a substantially larger exposure and will have greatly improved backgrounds because of overall better design and background control. These improvements are expected to enhance the SRDM sensitivity significantly, strengthening the role of skipper CCDs as low-threhold probes of sub-MeV hidden-sector dark matter.

\appendix
\section{Dielectric-response formalism}
\label{app:screening_formalism}

The dielectric screening factor used in the SRDM rate calculation is computed
from the inverse dielectric function of silicon,
\begin{equation}
    S_{\rm scr}(q,E_e)
    =
    \left|
        \epsilon^{-1}(q,E_e)
    \right|^2 .
    \label{eq:app_screening_factor}
\end{equation}
In the nominal signal model, \(\epsilon(q,E_e)\) is evaluated with the Mermin
prescription, following the dielectric-response formalism of
Ref.~\cite{Knapen_2022}. This model extends the Lindhard dielectric
function by introducing a finite damping parameter \(\Gamma\), which accounts
for dissipation in the material response.

The Mermin dielectric function is written as
\begin{equation}
    \epsilon_{\rm Mer}(q,E_e)
    =
    1+
    \frac{
        \left(1+i\Gamma/E_e\right)
        \left[
            \epsilon_{\rm Lind}(q,E_e+i\Gamma)-1
        \right]
    }{
        1+
        \left(i\Gamma/E_e\right)
        \frac{
            \epsilon_{\rm Lind}(q,E_e+i\Gamma)-1
        }{
            \epsilon_{\rm Lind}(q,0)-1
        }
    } .
    \label{eq:app_mermin}
\end{equation}
Here \(E_e\) is the electronic excitation energy entering the rate calculation.
In the limit \(\Gamma\to 0\), Eq.~\eqref{eq:app_mermin} reduces to the Lindhard
dielectric function. For finite \(\Gamma\), the plasmon pole is broadened and
the response includes dissipation while preserving local charge conservation.

The Lindhard dielectric function for an homogeneous electron gas is evaluated as
\begin{equation}
    \epsilon_{\rm Lind}(q,E_e)
    =
    1+
    \frac{3\omega_p^2}{q^2 v_F^2}\,
    f(u,z),
    \label{eq:app_lindhard}
\end{equation}
with
\begin{equation}
    u =
    \frac{E_e+i\eta}{qv_F},
    \qquad
    z =
    \frac{q}{2m_e v_F}.
    \label{eq:app_lindhard_uz}
\end{equation}
The auxiliary function \(f(u,z)\) is
\begin{equation}
    f(u,z)
    =
    \frac{1}{2}
    +
    \frac{1}{8z}
    \left[
        g(z-u)+g(z+u)
    \right],
    \label{eq:app_lindhard_f}
\end{equation}
where
\begin{equation}
    g(x)
    =
    \left(1-x^2\right)
    \log\left(
        \frac{1+x}{x-1}
    \right).
    \label{eq:app_lindhard_g}
\end{equation}
The infinitesimal positive parameter \(\eta\) fixes the retarded prescription of
the dielectric response. In the numerical implementation, the complex logarithm
branch is chosen consistently with this prescription. The explicit convention
in Eqs.~\eqref{eq:app_lindhard}--\eqref{eq:app_lindhard_g} was used because the
compact expressions in the literature are sensitive to sign and branch
conventions.

The Fermi velocity is obtained from the plasma frequency through the
homogeneous-electron-gas relation
\begin{equation}
    v_F
    =
    \left(
        \frac{3\pi \omega_p^2}
             {4\alpha m_e^2}
    \right)^{1/3},
    \label{eq:app_vf}
\end{equation}
using natural units. The plasma frequency is fixed by the silicon plasmon
energy, \(\omega_p=E_p=16.6~{\rm eV}\). The damping parameter is set to
\begin{equation}
    \Gamma = 3.2~{\rm eV},
    \label{eq:app_gamma}
\end{equation}
corresponding to the measured width of the silicon plasmon peak at small
momentum transfer~\cite{eels}.

The nominal dielectric function used in the rate calculation is therefore
\begin{equation}
    \epsilon(q,E_e)
    =
    \epsilon_{\rm Mer}(q,E_e),
    \label{eq:app_nominal_epsilon}
\end{equation}
and the screening factor entering Eq.~\eqref{eq:dR_dEe} is
\begin{equation}
    S_{\rm scr}(q,E_e)
    =
    \left|
        \epsilon_{\rm Mer}^{-1}(q,E_e)
    \right|^2 .
    \label{eq:app_nominal_screening}
\end{equation}

The implementation was validated by reproducing the reference dielectric-response
curves of Ref.~\cite{Knapen_2022}. The same implementation was then used to
evaluate the screening factor over the full \((q,E_e)\) phase space required by
the SRDM rate calculation.

\begin{figure}[tb]
    \centering
    \includegraphics[width=\linewidth]{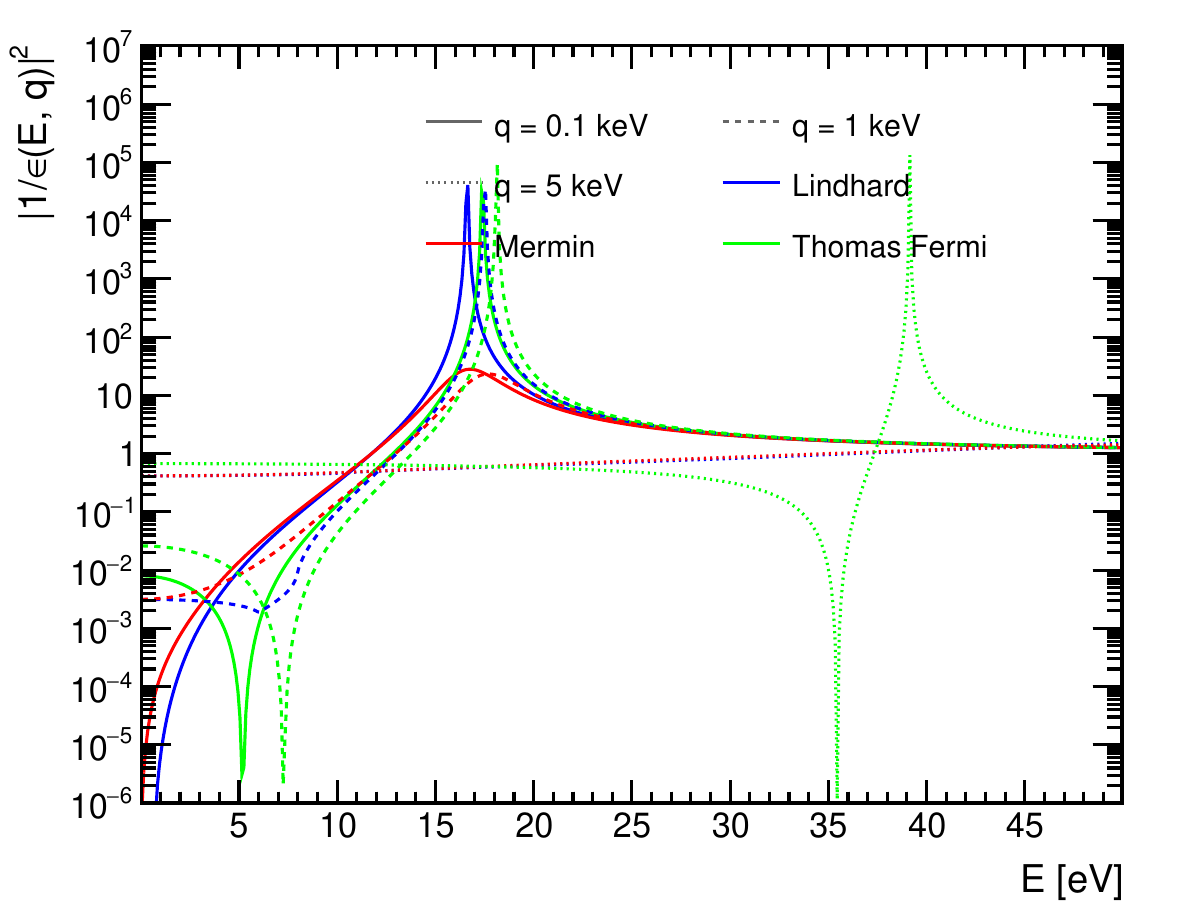}
    \caption{Screening factors as a function of energy for a set of momentum transfers. The nominal dielectric function used in the rate calculation with the Mermin formalism is presented against the original \qcdark{} function, the Thomas Fermi model, and the Lindhard model.}
    \label{fig:screening}
\end{figure}

\newpage
\bibliography{references.bib}

@article{Bertone:2004pz,
    author = "Bertone, Gianfranco and Hooper, Dan and Silk, Joseph",
    title = "{Particle dark matter: Evidence, candidates and constraints}",
    eprint = "hep-ph/0404175",
    archivePrefix = "arXiv",
    reportNumber = "FERMILAB-PUB-04-047-A",
    doi = "10.1016/j.physrep.2004.08.031",
    journal = "Phys. Rept.",
    volume = "405",
    pages = "279--390",
    year = "2005"
}

@article{Billard:2021uyg,
    author = "Billard, Julien and others",
    title = "{Direct detection of dark matter\textemdash{}APPEC committee report*}",
    eprint = "2104.07634",
    archivePrefix = "arXiv",
    primaryClass = "hep-ex",
    doi = "10.1088/1361-6633/ac5754",
    journal = "Rept. Prog. Phys.",
    volume = "85",
    number = "5",
    pages = "056201",
    year = "2022"
}

@article{Boveia:2018yeb,
    author = "Boveia, Antonio and Doglioni, Caterina",
    title = "{Dark Matter Searches at Colliders}",
    eprint = "1810.12238",
    archivePrefix = "arXiv",
    primaryClass = "hep-ex",
    doi = "10.1146/annurev-nucl-101917-021008",
    journal = "Ann. Rev. Nucl. Part. Sci.",
    volume = "68",
    pages = "429--459",
    year = "2018"
}

@article{An:2017ojc,
    author = "An, Haipeng and Pospelov, Maxim and Pradler, Josef and Ritz, Adam",
    title = "{Directly Detecting MeV-scale Dark Matter via Solar Reflection}",
    eprint = "1708.03642",
    archivePrefix = "arXiv",
    primaryClass = "hep-ph",
    doi = "10.1103/PhysRevLett.120.141801",
    journal = "Phys. Rev. Lett.",
    volume = "120",
    number = "14",
    pages = "141801",
    year = "2018",
    note = "[Erratum: Phys.Rev.Lett. 121, 259903 (2018)]"
}

@article{Boehm:2003ha,
    author = "Boehm, C. and Fayet, Pierre and Silk, J.",
    title = "{Light and heavy dark matter particles}",
    eprint = "hep-ph/0311143",
    archivePrefix = "arXiv",
    doi = "10.1103/PhysRevD.69.101302",
    journal = "Phys. Rev. D",
    volume = "69",
    pages = "101302",
    year = "2004"
}

@article{Drukier:1983gj,
    author = "Drukier, A. and Stodolsky, Leo",
    editor = "Tran Thanh Van, J.",
    title = "{Principles and Applications of a Neutral Current Detector for Neutrino Physics and Astronomy}",
    reportNumber = "MPI-PAE/PTh 36/82",
    doi = "10.1103/PhysRevD.30.2295",
    journal = "Phys. Rev. D",
    volume = "30",
    pages = "2295",
    year = "1984"
}

@article{CDEX:2023wfz,
    author = "Zhang, Z. Y. and others",
    collaboration = "CDEX",
    title = "{Experimental Limits on Solar Reflected Dark Matter with a New Approach on Accelerated-Dark-Matter{\textendash}Electron Analysis in Semiconductors}",
    eprint = "2309.14982",
    archivePrefix = "arXiv",
    primaryClass = "hep-ex",
    doi = "10.1103/PhysRevLett.132.171001",
    journal = "Phys. Rev. Lett.",
    volume = "132",
    number = "17",
    pages = "171001",
    year = "2024"
}

@misc{XENON:2025dpl,
    author = "Aprile, E. and others",
    collaboration = "XENON",
    title = "{Constraints on Solar Reflected Dark Matter from a combined analysis of XENON1T and XENONnT data}",
    eprint = "2512.19592",
    archivePrefix = "arXiv",
    primaryClass = "hep-ex",
    month = "12",
    year = "2025"
}

@article{Dreyer:2023ovn,
    author = "Dreyer, Cyrus E. and Essig, Rouven and Fernandez-Serra, Marivi and Singal, Aman and Zhen, Cheng",
    title = "{Fully ab-initio all-electron calculation of dark matter-electron scattering in crystals with evaluation of systematic uncertainties}",
    eprint = "2306.14944",
    archivePrefix = "arXiv",
    primaryClass = "hep-ph",
    doi = "10.1103/PhysRevD.109.115008",
    journal = "Phys. Rev. D",
    volume = "109",
    number = "11",
    pages = "115008",
    year = "2024"
}

@article{Kavanagh:2016pyr,
    author = "Kavanagh, Bradley J. and Catena, Riccardo and Kouvaris, Chris",
    title = "{Signatures of Earth-scattering in the direct detection of Dark Matter}",
    eprint = "1611.05453",
    archivePrefix = "arXiv",
    primaryClass = "hep-ph",
    reportNumber = "CP3-ORIGINS-2016-050",
    doi = "10.1088/1475-7516/2017/01/012",
    journal = "JCAP",
    volume = "01",
    pages = "012",
    year = "2017"
}

@article{Bringmann:2018cvk,
    author = "Bringmann, Torsten and Pospelov, Maxim",
    title = "{Novel direct detection constraints on light dark matter}",
    eprint = "1810.10543",
    archivePrefix = "arXiv",
    primaryClass = "hep-ph",
    doi = "10.1103/PhysRevLett.122.171801",
    journal = "Phys. Rev. Lett.",
    volume = "122",
    number = "17",
    pages = "171801",
    year = "2019"
}

@article{Ema:2018bih,
    author = "Ema, Yohei and Sala, Filippo and Sato, Ryosuke",
    title = "{Light Dark Matter at Neutrino Experiments}",
    eprint = "1811.00520",
    archivePrefix = "arXiv",
    primaryClass = "hep-ph",
    reportNumber = "DESY-18-194",
    doi = "10.1103/PhysRevLett.122.181802",
    journal = "Phys. Rev. Lett.",
    volume = "122",
    number = "18",
    pages = "181802",
    year = "2019"
}

@article{Wang:2021jic,
    author = "Wang, Jin-Wei and Granelli, Alessandro and Ullio, Piero",
    title = "{Direct Detection Constraints on Blazar-Boosted Dark Matter}",
    eprint = "2111.13644",
    archivePrefix = "arXiv",
    primaryClass = "astro-ph.HE",
    doi = "10.1103/PhysRevLett.128.221104",
    journal = "Phys. Rev. Lett.",
    volume = "128",
    number = "22",
    pages = "221104",
    year = "2022"
}

@article{Granelli:2022ysi,
    author = "Granelli, Alessandro and Ullio, Piero and Wang, Jin-Wei",
    title = "{Blazar-boosted dark matter at Super-Kamiokande}",
    eprint = "2202.07598",
    archivePrefix = "arXiv",
    primaryClass = "astro-ph.HE",
    reportNumber = "SISSA 02/2022/FISI",
    doi = "10.1088/1475-7516/2022/07/013",
    journal = "JCAP",
    volume = "07",
    number = "07",
    pages = "013",
    year = "2022"
}

@article{Cappiello:2022exa,
    author = "Cappiello, Christopher V. and Kozar, Neal P. Avis and Vincent, Aaron C.",
    title = "{Dark matter from Monogem}",
    eprint = "2210.09448",
    archivePrefix = "arXiv",
    primaryClass = "hep-ph",
    doi = "10.1103/PhysRevD.107.035003",
    journal = "Phys. Rev. D",
    volume = "107",
    number = "3",
    pages = "035003",
    year = "2023"
}

@misc{Lantero-Barreda:2025sgy,
    author = "Lantero-Barreda, Agust{\'\i}n and Centeno-Lorca, Carlos and Kavanagh, Bradley J. and Castello-Mor, N{\'u}ria",
    title = "{A Fast Earth-scattering Formalism for Light Dark Matter with Dark Photon Mediators}",
    eprint = "2511.10589",
    archivePrefix = "arXiv",
    primaryClass = "hep-ph",
    month = "11",
    year = "2025"
}

@misc{DAMIC-M:2025ltz,
    author = "Aggarwal, K. and others",
    collaboration = "DAMIC-M",
    title = "{Daily Modulation Constraints on Light Dark Matter with DAMIC-M}",
    eprint = "2511.13962",
    archivePrefix = "arXiv",
    primaryClass = "hep-ex",
    month = "11",
    year = "2025"
}

@article{lindhard1954,
  title = {On the properties of a gas of charged particles},
  author = {Lindhard, Jens},
  journal = {Danske Matematisk-fysiske Meddelelser},
  volume = {28},
  number = {8},
  pages = {1--57},
  year = {1954},
  publisher = {Munksgaard}
}

@misc{Dreyer:2026bmz,
    author = "Dreyer, Cyrus and Essig, Rouven and Fernandez-Serra, Marivi and Hott, Megan and Singal, Aman",
    title = "{All-electron dark matter-electron scattering with random-phase approximation dielectric screening and local field effects}",
    eprint = "2603.12326",
    archivePrefix = "arXiv",
    primaryClass = "hep-ph",
    month = "3",
    year = "2026"
}

@article{Mermin:1970zz,
    author = "Mermin, N. D.",
    title = "{Lindhard Dielectric Function in the Relaxation-Time Approximation}",
    doi = "10.1103/PhysRevB.1.2362",
    journal = "Phys. Rev. B",
    volume = "1",
    pages = "2362--2363",
    year = "1970"
}

@misc{Cerdeno:2010jj,
    author = "Cerdeno, David G. and Green, Anne M.",
    title = "{Direct detection of WIMPs}",
    eprint = "1002.1912",
    archivePrefix = "arXiv",
    primaryClass = "astro-ph.CO",
    reportNumber = "FTUAM-10-02, IFT-UAM-CSIC-10-07, FTUAM-10-02; IFT-UAM/CSIC-10-07",
    doi = "10.1017/CBO9780511770739.018",
    pages = "347--369",
    month = "2",
    year = "2010"
}

@article{Green:2017odb,
    author = "Green, Anne M",
    title = "{Astrophysical uncertainties on the local dark matter distribution and direct detection experiments}",
    eprint = "1703.10102",
    archivePrefix = "arXiv",
    primaryClass = "astro-ph.CO",
    doi = "10.1088/1361-6471/aa7819",
    journal = "J. Phys. G",
    volume = "44",
    number = "8",
    pages = "084001",
    year = "2017"
}

@article{DAMIC-M:2023gxo,
    author = "Arnquist, I. and others",
    collaboration = "DAMIC-M",
    title = "{First Constraints from DAMIC-M on Sub-GeV Dark-Matter Particles Interacting with Electrons}",
    eprint = "2302.02372",
    archivePrefix = "arXiv",
    primaryClass = "hep-ex",
    reportNumber = "FERMILAB-PUB-23-039-PPD",
    doi = "10.1103/PhysRevLett.130.171003",
    journal = "Phys. Rev. Lett.",
    volume = "130",
    number = "17",
    pages = "171003",
    year = "2023"
}

@article{SENSEI:2023zdf,
    author = "Adari, Prakruth and others",
    collaboration = "SENSEI",
    title = "{First Direct-Detection Results on Sub-GeV Dark Matter Using the SENSEI Detector at SNOLAB}",
    eprint = "2312.13342",
    archivePrefix = "arXiv",
    primaryClass = "astro-ph.CO",
    reportNumber = "YITP-SB-2023-30, FERMILAB-PUB-23-0824-CSAID-PPD",
    doi = "10.1103/PhysRevLett.134.011804",
    journal = "Phys. Rev. Lett.",
    volume = "134",
    number = "1",
    pages = "011804",
    year = "2025"
}

@article{DAMIC:2011khz,
    author = "Barreto, J. and others",
    collaboration = "DAMIC",
    title = "{Direct Search for Low Mass Dark Matter Particles with CCDs}",
    eprint = "1105.5191",
    archivePrefix = "arXiv",
    primaryClass = "astro-ph.IM",
    reportNumber = "FERMILAB-PUB-11-289-AE",
    doi = "10.1016/j.physletb.2012.04.006",
    journal = "Phys. Lett. B",
    volume = "711",
    pages = "264--269",
    year = "2012"
}

@article{DAMICsnolab,
  title = "{Search for low-mass WIMPs in a 0.6 kg day exposure of the DAMIC experiment at SNOLAB}",
  author = {Aguilar-Arevalo, A. and others},
  collaboration = {DAMIC Collaboration},
  journal = {Phys. Rev. D},
  volume = {94},
  issue = {8},
  pages = {082006},
  numpages = {11},
  year = {2016},
  month = {Oct},
  publisher = {American Physical Society},
  doi = {10.1103/PhysRevD.94.082006},
  url = {https://link.aps.org/doi/10.1103/PhysRevD.94.082006}
}

@article{Goodman:1984dc,
    author = "Goodman, Mark W. and Witten, Edward",
    editor = "Srednicki, M. A.",
    title = "{Detectability of Certain Dark Matter Candidates}",
    reportNumber = "Print-85-0030 (PRINCETON)",
    doi = "10.1103/PhysRevD.31.3059",
    journal = "Phys. Rev. D",
    volume = "31",
    pages = "3059",
    year = "1985"
}

@article{Drukier:1986tm,
    author = "Drukier, A. K. and Freese, Katherine and Spergel, D. N.",
    title = "{Detecting Cold Dark Matter Candidates}",
    doi = "10.1103/PhysRevD.33.3495",
    journal = "Phys. Rev. D",
    volume = "33",
    pages = "3495--3508",
    year = "1986"
}

@article{Pospelov:2007mp,
    author = "Pospelov, Maxim and Ritz, Adam and Voloshin, Mikhail B.",
    title = "{Secluded WIMP Dark Matter}",
    eprint = "0711.4866",
    archivePrefix = "arXiv",
    primaryClass = "hep-ph",
    doi = "10.1016/j.physletb.2008.02.052",
    journal = "Phys. Lett. B",
    volume = "662",
    pages = "53--61",
    year = "2008"
}

@article{Hooper:2008im,
    author = "Hooper, Dan and Zurek, Kathryn M.",
    title = "{A Natural Supersymmetric Model with MeV Dark Matter}",
    eprint = "0801.3686",
    archivePrefix = "arXiv",
    primaryClass = "hep-ph",
    reportNumber = "FERMILAB-PUB-07-587-A",
    doi = "10.1103/PhysRevD.77.087302",
    journal = "Phys. Rev. D",
    volume = "77",
    pages = "087302",
    year = "2008"
}

@article{Chu:2011be,
    author = "Chu, Xiaoyong and Hambye, Thomas and Tytgat, Michel H. G.",
    title = "{The Four Basic Ways of Creating Dark Matter Through a Portal}",
    eprint = "1112.0493",
    archivePrefix = "arXiv",
    primaryClass = "hep-ph",
    reportNumber = "ULB-TH-11-26",
    doi = "10.1088/1475-7516/2012/05/034",
    journal = "JCAP",
    volume = "05",
    pages = "034",
    year = "2012"
}

@article{Knapen:2017xzo,
    author = "Knapen, Simon and Lin, Tongyan and Zurek, Kathryn M.",
    title = "{Light Dark Matter: Models and Constraints}",
    eprint = "1709.07882",
    archivePrefix = "arXiv",
    primaryClass = "hep-ph",
    doi = "10.1103/PhysRevD.96.115021",
    journal = "Phys. Rev. D",
    volume = "96",
    number = "11",
    pages = "115021",
    year = "2017"
}

@article{Emken:2017hnp,
    author = "Emken, Timon and Kouvaris, Chris and Nielsen, Niklas Gr{\o}nlund",
    title = "{The Sun as a sub-GeV Dark Matter Accelerator}",
    eprint = "1709.06573",
    archivePrefix = "arXiv",
    primaryClass = "hep-ph",
    reportNumber = "CP3-ORIGINS-2017-035",
    doi = "10.1103/PhysRevD.97.063007",
    journal = "Phys. Rev. D",
    volume = "97",
    number = "6",
    pages = "063007",
    year = "2018"
}

@article{Essig:2015cda,
    author = "Essig, Rouven and Fernandez-Serra, Marivi and Mardon, Jeremy and Soto, Adrian and Volansky, Tomer and Yu, Tien-Tien",
    title = "{Direct Detection of sub-GeV Dark Matter with Semiconductor Targets}",
    eprint = "1509.01598",
    archivePrefix = "arXiv",
    primaryClass = "hep-ph",
    doi = "10.1007/JHEP05(2016)046",
    journal = "JHEP",
    volume = "05",
    pages = "046",
    year = "2016"
}

@article{Gaskins:2016cha,
    author = "Gaskins, Jennifer M.",
    title = "{A review of indirect searches for particle dark matter}",
    eprint = "1604.00014",
    archivePrefix = "arXiv",
    primaryClass = "astro-ph.HE",
    doi = "10.1080/00107514.2016.1175160",
    journal = "Contemp. Phys.",
    volume = "57",
    number = "4",
    pages = "496--525",
    year = "2016"
}

@article{Essig:2011nj,
    author = "Essig, Rouven and Mardon, Jeremy and Volansky, Tomer",
    title = "{Direct Detection of Sub-GeV Dark Matter}",
    eprint = "1108.5383",
    archivePrefix = "arXiv",
    primaryClass = "hep-ph",
    reportNumber = "SLAC-PUB-14538",
    doi = "10.1103/PhysRevD.85.076007",
    journal = "Phys. Rev. D",
    volume = "85",
    pages = "076007",
    year = "2012"
}

@article{Kouvaris:2014lpa,
    author = "Kouvaris, Chris and Shoemaker, Ian M.",
    title = "{Daily modulation as a smoking gun of dark matter with significant stopping rate}",
    eprint = "1405.1729",
    archivePrefix = "arXiv",
    primaryClass = "hep-ph",
    reportNumber = "CP3-ORIGINS-2014-019, DIAS-2014-19",
    doi = "10.1103/PhysRevD.90.095011",
    journal = "Phys. Rev. D",
    volume = "90",
    pages = "095011",
    year = "2014"
}

@article{DaMaSCUS-SUN:2024,
    author = "Emken, Timon and Essig, Rouven and Xu, Hailin",
    title = "{Solar reflection of dark matter with dark-photon mediators}",
    eprint = "2404.10066",
    archivePrefix = "arXiv",
    primaryClass = "hep-ph",
    doi = "10.1088/1475-7516/2024/07/023",
    journal = "JCAP",
    volume = "07",
    pages = "023",
    year = "2024"
}

@article{DaMaSCUS-SUN:2022,
  title = {Solar reflection of light dark matter with heavy mediators},
  author = {Emken, Timon},
  journal = {Phys. Rev. D},
  volume = {105},
  issue = {6},
  pages = {063020},
  numpages = {35},
  year = {2022},
  month = {Mar},
  publisher = {American Physical Society},
  doi = {10.1103/PhysRevD.105.063020},
  url = {https://link.aps.org/doi/10.1103/PhysRevD.105.063020}
}

@article{Cowan:2013,
author = "Cowan, Glen and Cranmer, Kyle and Gross, Eilam and Vitells, Ofer",
    title = "{Asymptotic formulae for likelihood-based tests of new physics}",
    eprint = "1007.1727",
    archivePrefix = "arXiv",
    primaryClass = "physics.data-an",
    doi = "10.1140/epjc/s10052-011-1554-0",
    journal = "Eur. Phys. J. C",
    volume = "71",
    pages = "1554",
    year = "2011",
    note = "[Erratum: Eur.Phys.J.C 73, 2501 (2013)]"
}

@article{Baxter:2021pqo,
    author = "Baxter, D. and others",
    title = "{Recommended conventions for reporting results from direct dark matter searches}",
    eprint = "2105.00599",
    archivePrefix = "arXiv",
    primaryClass = "hep-ex",
    doi = "10.1140/epjc/s10052-021-09655-y",
    journal = "Eur. Phys. J. C",
    volume = "81",
    number = "10",
    pages = "907",
    year = "2021"
}

@misc{Verne2,
      title={A Fast Earth-scattering Formalism for Light Dark Matter with Dark Photon Mediators}, 
      author={Agustín Lantero-Barreda and Carlos Centeno and Bradley J. Kavanagh and Nuria Castelló-Mor},
      year={2025},
      eprint={2511.10589},
      archivePrefix={arXiv},
      primaryClass={hep-ph},
      url={https://arxiv.org/abs/2511.10589}, 
}

@article{Ramanathan:2020,
  title = "{Ionization yield in silicon for eV-scale electron-recoil processes}",
  author = {Ramanathan, K. and Kurinsky, N.},
  journal = {Phys. Rev. D},
  volume = {102},
  issue = {6},
  pages = {063026},
  numpages = {14},
  year = {2020},
  month = {Sep},
  publisher = {American Physical Society},
  doi = {10.1103/PhysRevD.102.063026},
  url = {https://link.aps.org/doi/10.1103/PhysRevD.102.063026}
}

@article{DAMIC-M:2023DM,
    author = "Arnquist, I. and others",
    collaboration = "DAMIC-M",
    title = "{Search for Daily Modulation of MeV Dark Matter Signals with DAMIC-M}",
    eprint = "2307.07251",
    archivePrefix = "arXiv",
    primaryClass = "hep-ex",
    doi = "10.1103/PhysRevLett.132.101006",
    journal = "Phys. Rev. Lett.",
    volume = "132",
    number = "10",
    pages = "101006",
    year = "2024"
}

@article{DAMIC-M:2025,
  title = "{Probing Benchmark Models of Hidden-Sector Dark Matter with DAMIC-M}",
  author = {Aggarwal, K. and Arnquist, I. and Avalos, N. and Bertou, X. and Castell\'o-Mor, N. and Chavarria, A. E. and Cuevas-Zepeda, J. and Dastgheibi-Fard, A. and De Dominicis, C. and Deligny, O. and Duarte-Campderros, J. and Estrada, E. and Ga\"{\i}or, R. and Gkougkousis, E.-L. and Hossbach, T. and Iddir, L. and Kavanagh, B. J. and Kilminster, B. and Lantero-Barreda, A. and Lawson, I. and Letessier-Selvon, A. and Lin, H. and Loaiza, P. and Lopez-Virto, A. and Lou, R. and McGuire, K. J. and Munagavalasa, S. and Noonan, J. and Norcini, D. and Paul, S. and Privitera, P. and Robmann, P. and Roach, B. and Settimo, M. and Smida, R. and Traina, M. and Vilar, R. and Yajur, R. and Venegas-Vargas, D. and Zhu, C. and Zhu, Y.},
  collaboration = {DAMIC-M Collaboration},
  journal = {Phys. Rev. Lett.},
  volume = {135},
  issue = {7},
  pages = {071002},
  numpages = {10},
  year = {2025},
  month = {Aug},
  publisher = {American Physical Society},
  doi = {10.1103/2tcc-bqck},
  url = {https://link.aps.org/doi/10.1103/2tcc-bqck}
}

@article{LBC:2024,
doi = {10.1088/1748-0221/19/11/T11010},
url = {https://dx.doi.org/10.1088/1748-0221/19/11/T11010},
year = {2024},
month = {nov},
publisher = {IOP Publishing},
volume = {19},
number = {11},
pages = {T11010},
author = {Arnquist, I. and Avalos, N. and Bailly, P. and Baxter, D. and Bertou, X. and Bogdan, M. and Bourgeois, C. and Brandt, J. and Cadiou, A. and Castelló-Mor, N. and Chavarria, A.E. and Conde, M. and Cuevas-Zepeda, J. and Dastgheibi-Fard, A. and De Dominicis, C. and Deligny, O. and Desani, R. and Dhellot, M. and Duarte-Campderros, J. and Estrada, E. and Florin, D. and Gadola, N. and Gaïor, R. and Gkougkousis, E.-L. and González Sánchez, J. and Hope, S. and Hossbach, T. and Huehn, M. and Kallander, M. and Kilminster, B. and Iddir, L. and Lantero-Barreda, A. and Lawson, I. and Lebbolo, H. and Lee, S. and Leray, P. and Letessier Selvon, A. and Lin, H. and Loaiza, P. and Lopez-Virto, A. and Martin, D. and McGuire, K.J. and Milleto, T. and Mitra, P. and Moya Martin, D. and Munagavalasa, S. and Norcini, D. and Overman, C. and Paul, S. and Peterson, D. and Piers, A. and Pochon, O. and Privitera, P. and Reynet, D. and Roach, B.A. and Robmann, P. and Roehnelt, R. and Settimo, M. and Smee, S. and Smida, R. and Stillwell, B. and Van Wechel, T. and Traina, M. and Vilar, R. and Vollhardt, A. and Warot, G. and Wolf, D. and Yajur, R. and Zopounidis, J-P. and The DAMIC-M collaboration},
title = "{The DAMIC-M Low Background Chamber}",
journal = {Journal of Instrumentation}
}

@article{eels,
author = {Egerton, Ray},
year = {2009},
month = {01},
pages = {16502-25},
title = {Electron Energy Loss Spectroscopy in the TEM},
volume = {72},
journal = {REPORTS ON PROGRESS IN PHYSICS Rep. Prog. Phys},
doi = {10.1088/0034-4885/72/1/016502}
}

@article{Knapen_2022,
   title={python package for dark matter scattering in dielectric targets},
   volume={105},
   ISSN={2470-0029},
   url={http://dx.doi.org/10.1103/PhysRevD.105.015014},
   DOI={10.1103/physrevd.105.015014},
   number={1},
   journal={Physical Review D},
   publisher={American Physical Society (APS)},
   author={Knapen, Simon and Kozaczuk, Jonathan and Lin, Tongyan},
   year={2022},
   month=Jan }

@misc{xenoncollaboration2025constraintssolarreflecteddark,
      title={Constraints on Solar Reflected Dark Matter from a combined analysis of XENON1T and XENONnT data}, 
      author={XENON Collaboration and E. Aprile and J. Aalbers and K. Abe and M. Adrover and S. Ahmed Maouloud and L. Althueser and B. Andrieu and E. Angelino and D. Ant'on Martin and S. R. Armbruster and F. Arneodo and L. Baudis and M. Bazyk and L. Bellagamba and R. Biondi and A. Bismark and K. Boese and R. M. Braun and G. Bruni and G. Bruno and R. Budnik and C. Cai and C. Capelli and J. M. R. Cardoso and A. P. Cimental Ch'avez and A. P. Colijn and J. Conrad and J. J. Cuenca-García and V. D'Andrea and L. C. Daniel Garcia and M. P. Decowski and A. Deisting and C. Di Donato and P. Di Gangi and S. Diglio and K. Eitel and S. el Morabit and R. Elleboro and A. Elykov and A. D. Ferella and C. Ferrari and H. Fischer and T. Flehmke and M. Flierman and R. Frankel and D. Fuchs and W. Fulgione and C. Fuselli and R. Gaior and F. Gao and R. Giacomobono and F. Girard and R. Glade-Beucke and L. Grandi and J. Grigat and H. Guan and M. Guida and P. Gyorgy and R. Hammann and A. Higuera and C. Hils and L. Hoetzsch and N. F. Hood and M. Iacovacci and Y. Itow and J. Jakob and F. Joerg and Y. Kaminaga and M. Kara and S. Kazama and P. Kharbanda and M. Kobayashi and D. Koke and K. Kooshkjalali and A. Kopec and H. Landsman and R. F. Lang and L. Levinson and I. Li and S. Li and S. Liang and Z. Liang and Y. -T. Lin and S. Lindemann and M. Lindner and K. Liu and M. Liu and J. Loizeau and F. Lombardi and J. A. M. Lopes and G. M. Lucchetti and T. Luce and Y. Ma and C. Macolino and J. Mahlstedt and F. Marignetti and T. Marrod'an Undagoitia and K. Martens and J. Masbou and S. Mastroianni and V. Mazza and A. Melchiorre and J. Merz and M. Messina and A. J. P. Michel and K. Miuchi and A. Molinario and S. Moriyama and K. Morå and M. Murra and J. Müller and K. Ni and C. T. Oba Ishikawa and U. Oberlack and S. Ouahada and B. Paetsch and Y. Pan and Q. Pellegrini and R. Peres and J. Pienaar and M. Pierre and G. Plante and T. R. Pollmann and A. Prajapati and L. Principe and J. Qin and D. Ram'irez Garcia and A. Ravindran and A. Razeto and R. Singh and L. Sanchez and J. M. F. dos Santos and I. Sarnoff and G. Sartorelli and J. Schreiner and P. Schulte and H. Schulze Eißing and M. Schumann and L. Scotto Lavina and M. Selvi and F. Semeria and F. N. Semler and P. Shagin and S. Shi and H. Simgen and Z. Song and A. Stevens and C. Szyszka and A. Takeda and Y. Takeuchi and P. -L. Tan and D. Thers and G. Trinchero and C. D. Tunnell and K. Valerius and S. Vecchi and S. Vetter and G. Volta and C. Weinheimer and M. Weiss and D. Wenz and C. Wittweg and V. H. S. Wu and Y. Xing and D. Xu and Z. Xu and M. Yamashita and J. Yang and L. Yang and J. Ye and M. Yoshida and L. Yuan and G. Zavattini and Y. Zhao and M. Zhong and T. Zhu},
      year={2025},
      eprint={2512.19592},
      archivePrefix={arXiv},
      primaryClass={hep-ex},
      url={https://arxiv.org/abs/2512.19592}, 
}

@misc{freezein,
      title={US Cosmic Visions: New Ideas in Dark Matter 2017: Community Report}, 
      author={Marco Battaglieri and Alberto Belloni and Aaron Chou and Priscilla Cushman and Bertrand Echenard and Rouven Essig and Juan Estrada and Jonathan L. Feng and Brenna Flaugher and Patrick J. Fox and Peter Graham and Carter Hall and Roni Harnik and JoAnne Hewett and Joseph Incandela and Eder Izaguirre and Daniel McKinsey and Matthew Pyle and Natalie Roe and Gray Rybka and Pierre Sikivie and Tim M. P. Tait and Natalia Toro and Richard Van De Water and Neal Weiner and Kathryn Zurek and Eric Adelberger and Andrei Afanasev and Derbin Alexander and James Alexander and Vasile Cristian Antochi and David Mark Asner and Howard Baer and Dipanwita Banerjee and Elisabetta Baracchini and Phillip Barbeau and Joshua Barrow and Noemie Bastidon and James Battat and Stephen Benson and Asher Berlin and Mark Bird and Nikita Blinov and Kimberly K. Boddy and Mariangela Bondi and Walter M. Bonivento and Mark Boulay and James Boyce and Maxime Brodeur and Leah Broussard and Ranny Budnik and Philip Bunting and Marc Caffee and Sabato Stefano Caiazza and Sheldon Campbell and Tongtong Cao and Gianpaolo Carosi and Massimo Carpinelli and Gianluca Cavoto and Andrea Celentano and Jae Hyeok Chang and Swapan Chattopadhyay and Alvaro Chavarria and Chien-Yi Chen and Kenneth Clark and John Clarke and Owen Colegrove and Jonathon Coleman and David Cooke and Robert Cooper and Michael Crisler and Paolo Crivelli and Francesco D'Eramo and Domenico D'Urso and Eric Dahl and William Dawson and Marzio De Napoli and Raffaella De Vita and Patrick DeNiverville and Stephen Derenzo and Antonia Di Crescenzo and Emanuele Di Marco and Keith R. Dienes and Milind Diwan and Dongwi Handiipondola Dongwi and Alex Drlica-Wagner and Sebastian Ellis and Anthony Chigbo Ezeribe and Glennys Farrar and Francesc Ferrer and Enectali Figueroa-Feliciano and Alessandra Filippi and Giuliana Fiorillo and Bartosz Fornal and Arne Freyberger and Claudia Frugiuele and Cristian Galbiati and Iftah Galon and Susan Gardner and Andrew Geraci and Gilles Gerbier and Mathew Graham and Edda Gschwendtner and Christopher Hearty and Jaret Heise and Reyco Henning and Richard J. Hill and David Hitlin and Yonit Hochberg and Jason Hogan and Maurik Holtrop and Ziqing Hong and Todd Hossbach and T. B. Humensky and Philip Ilten and Kent Irwin and John Jaros and Robert Johnson and Matthew Jones and Yonatan Kahn and Narbe Kalantarians and Manoj Kaplinghat and Rakshya Khatiwada and Simon Knapen and Michael Kohl and Chris Kouvaris and Jonathan Kozaczuk and Gordan Krnjaic and Valery Kubarovsky and Eric Kuflik and Alexander Kusenko and Rafael Lang and Kyle Leach and Tongyan Lin and Mariangela Lisanti and Jing Liu and Kun Liu and Ming Liu and Dinesh Loomba and Joseph Lykken and Katherine Mack and Jeremiah Mans and Humphrey Maris and Thomas Markiewicz and Luca Marsicano and C. J. Martoff and Giovanni Mazzitelli and Christopher McCabe and Samuel D. McDermott and Art McDonald and Bryan McKinnon and Dongming Mei and Tom Melia and Gerald A. Miller and Kentaro Miuchi and Sahara Mohammed Prem Nazeer and Omar Moreno and Vasiliy Morozov and Frederic Mouton and Holger Mueller and Alexander Murphy and Russell Neilson and Tim Nelson and Christopher Neu and Yuri Nosochkov and Ciaran O'Hare and Noah Oblath and John Orrell and Jonathan Ouellet and Saori Pastore and Sebouh Paul and Maxim Perelstein and Annika Peter and Nguyen Phan and Nan Phinney and Michael Pivovaroff and Andrea Pocar and Maxim Pospelov and Josef Pradler and Paolo Privitera and Stefano Profumo and Mauro Raggi and Surjeet Rajendran and Nunzio Randazzo and Tor Raubenheimer and Christian Regenfus and Andrew Renshaw and Adam Ritz and Thomas Rizzo and Leslie Rosenberg and Andre Rubbia and Ben Rybolt and Tarek Saab and Benjamin R. Safdi and Elena Santopinto and Andrew Scarff and Michael Schneider and Philip Schuster and George Seidel and Hiroyuki Sekiya and Ilsoo Seong and Gabriele Simi and Valeria Sipala and Tracy Slatyer and Oren Slone and Peter F Smith and Jordan Smolinsky and Daniel Snowden-Ifft and Matthew Solt and Andrew Sonnenschein and Peter Sorensen and Neil Spooner and Brijesh Srivastava and Ion Stancu and Louis Strigari and Jan Strube and Alexander O. Sushkov and Matthew Szydagis and Philip Tanedo and David Tanner and Rex Tayloe and William Terrano and Jesse Thaler and Brooks Thomas and Brianna Thorpe and Thomas Thorpe and Javier Tiffenberg and Nhan Tran and Marco Trovato and Christopher Tully and Tony Tyson and Tanmay Vachaspati and Sven Vahsen and Karl van Bibber and Justin Vandenbroucke and Anthony Villano and Tomer Volansky and Guojian Wang and Thomas Ward and William Wester and Andrew Whitbeck and David A. Williams and Matthew Wing and Lindley Winslow and Bogdan Wojtsekhowski and Hai-Bo Yu and Shin-Shan Yu and Tien-Tien Yu and Xilin Zhang and Yue Zhao and Yi-Ming Zhong},
      year={2017},
      eprint={1707.04591},
      archivePrefix={arXiv},
      primaryClass={hep-ph},
      url={https://arxiv.org/abs/1707.04591}, 
}

@article{stellar_constrains,
   author = {Chang, Jae Hyeok and Essig, Rouven and Reinert, Annika},
   title = "{Light(ly)-coupled dark matter in the keV range: freeze-in and constraints}",
   journal = "JHEP",
   volume = "03",
   pages = "141",
   year = "2021",
   doi = "10.1007/JHEP03(2021)141"
}

\end{document}